\pdfoutput=1

	\def\clsstyle{prb} 
	\newcommand{\sectionheading}[1]{\section{#1}}

	\newcommand{\prltext}[1]{}
	
\documentclass[aps,\clsstyle,reprint,twocolumn,showpacs,superscriptaddress]{revtex4-1}

\usepackage{amsmath, amssymb, bm, braket, dsfont, times, mathtools,graphicx,color}
\usepackage[breaklinks,colorlinks,allcolors=blue]{hyperref}

\usepackage[normalem]{ulem}

\newcommand*{\Scale}[2][4]{\scalebox{#1}{$#2$}}%

\begin{document}

\title{One-dimensional symmetry protected topological phases and their transitions}
\author{Ruben Verresen}
\affiliation{Department of Physics, T42, Technische Universit\"at M\"unchen, 85748 Garching, Germany}
\affiliation{Max-Planck-Institute for the Physics of Complex Systems, 01187 Dresden, Germany}
\author{Roderich Moessner}
\affiliation{Max-Planck-Institute for the Physics of Complex Systems, 01187 Dresden, Germany}
\author{Frank Pollmann}
\affiliation{Department of Physics, T42, Technische Universit\"at M\"unchen, 85748 Garching, Germany}
\affiliation{Max-Planck-Institute for the Physics of Complex Systems, 01187 Dresden, Germany}
\date{\today}
\begin{abstract}
We present a unified perspective on symmetry protected topological (SPT) phases in one dimension and address the open question of what characterizes their phase transitions. In the first part of this work we use symmetry as a guide to map various well-known fermionic and spin SPTs to a Kitaev chain with coupling of range $\alpha \in \mathbb Z$.
%This unified picture uncovers new properties of old models --such as how the cluster state is the fixed point limit of the Affleck-Kennedy-Lieb-Tasaki state in disguise-- and leads to new SPTs --such as the Hubbard chain interpolating between the stack of four Kitaev chains and a spin chain in the Haldane phase.
This unified picture uncovers new properties of old models --such as how the cluster state is the fixed point limit of the Affleck-Kennedy-Lieb-Tasaki state in disguise-- and elucidates the connection between fermionic and bosonic phases --with the Hubbard chain interpolating between four Kitaev chains and a spin chain in the Haldane phase.
In the second part, we study the topological phase transitions between these models in the presence of interactions. This leads us to conjecture that the critical point between any SPT with $d$-dimensional edge modes and the trivial phase has a central charge $c \geq \log_2 d$. We analytically verify this for many known transitions. This agrees with the intuitive notion that the phase transition is described by a delocalized edge mode, and that the central charge of a conformal field theory is a measure of the gapless degrees of freedom.
\end{abstract}

\maketitle

Topology has established itself as a fundamental principle in condensed matter physics. For gapped ground states of local Hamiltonians, topological invariants can label distinct phases of matter, and these non-local order parameters can be associated with exotic features such as protected edge states or anyonic excitations\cite{Wen16}. While the classification of topological phases has been achieved for non-interacting fermions in arbitrary dimensions\cite{Fu2007,Kitaev09,Schnyder-2008,Altland97,Ryu10}, the extension to systems of interacting particles is a matter of ongoing work. For gapped systems in one spatial dimension, however, the general principles have been elucidated\cite{Pollmann10,Fidkowski11,Turner-2010,Chen-2010,Schuch-2011,SPt}. In particular it is known that topological invariants require the presence of an unbroken symmetry in order to be well-defined. These label so-called \emph{symmetry protected topological (SPT) phases}.

One-dimensional SPT phases have the curious property that the physical edges have modes at zero energy. These are protected by how particular bulk symmetries act anomalously on the edge\cite{Pollmann10,Chen-2010,Schuch-2011,Fidkowski11,Turner-2010} (in section \ref{scn:SPT} and Appendix \ref{app:sym} we present an accessible review of the classification of one-dimensional SPT phases). An archetype is the Haldane phase, realized by the spin-$1$ Heisenberg chain with its spin-$\frac{1}{2}$ edge modes: the bulk is symmetric with respect to $SO(3)$ whereas the edges transform under $SU(2)$. While that particular model is not analytically tractable, there are a number of exactly soluble fermionic and spin chains that have been uncovered over the decades realizing SPT phases. One might wonder whether there are links between these distinct models. This is the question we address in the first part of this work, leading to a unification of various models by relating them to stacks of Kitaev chains\cite{Kitaev-2001}.

This unified set of models provides a framework for our second topic: ``What characterizes the critical theory between SPT phases?''. There have been various works studying the transitions between particular SPT phases\cite{Jiang10,Liu11,Son11,Nonne13,Ueda14,Morimoto14,Chen15,Prakash16,Rao16,SON,Ohta16}, but it has proven difficult to make quantitative statements about the general case\cite{Chen13,Tsui15,Tsui17}. The latter works have led to the intuitive picture that the gapless fields at the transition are in some sense the delocalized boundary excitations. Our goal is to quantify this intuition by establishing a relationship between the number of low-energy degrees of freedom at the transition and the number of edge modes in the neighboring gapped phases.

The main outcome of the first part of this paper is that various SPT models can be related to stacks of Kitaev chains. The Kitaev chain has a single Majorana zero mode on each edge, but by stacking multiple copies one can have an arbitrary number of such modes. In the classification of non-interacting SPT phases (i.e. topological insulators and superconductors)\cite{Altland97,Schnyder-2008,Kitaev09}, spinless time reversal symmetry (TRS) prevents these Majorana modes from gapping out. Such stacks of Kitaev chains were an important testing ground to subsequently uncover the classification of \emph{interacting} SPTs. In the presence of interactions, there are only eight distinct phases protected by TRS, characterized by how fermionic parity symmetry and TRS are represented on the edge\cite{Fidkowski11,Turner-2010}. Here we revisit these stacks. As explained in the main text, for every $\alpha \in \mathbb Z$, a stack of $\alpha$ Kitaev chains 
($\alpha < 0$ denoting spatially inverted chains) is equivalent to a single Kitaev chain with coupling of range $\alpha$. With economy of language, we refer to this as the $\alpha$-chain.

\begin{figure}
	\includegraphics[scale=.43]{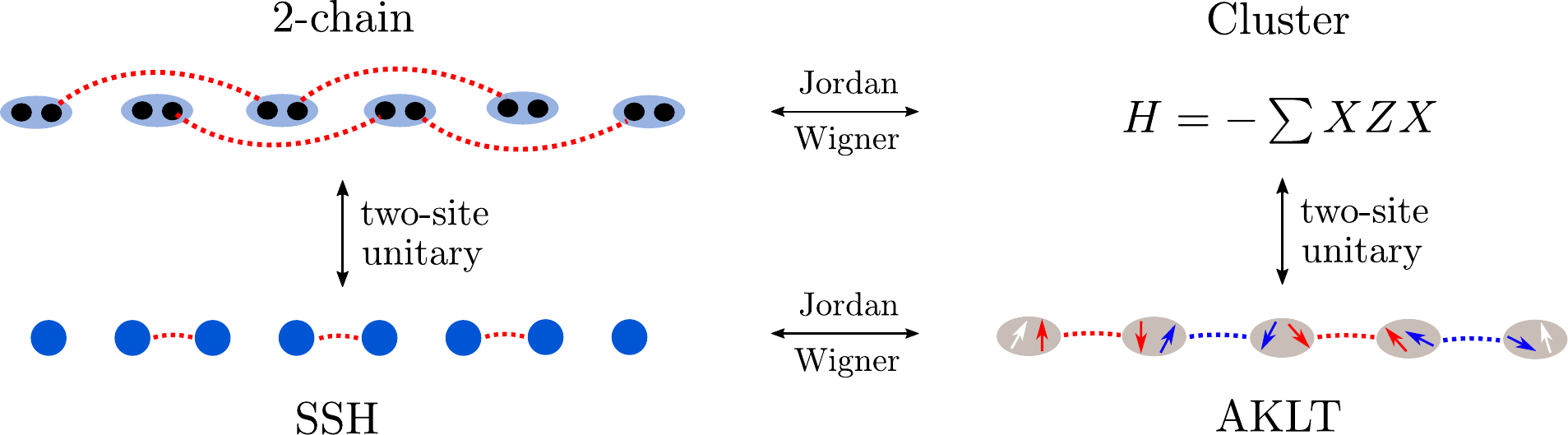}
	\caption{SPT models related to the $\alpha$-chain with $\alpha = 2$. In the case of the AKLT model it is at the level of the ground state, whereas for the other three models it is at the level of the full Hamiltonian.} \label{fig:equiv}
\end{figure}

Let us highlight a few of our findings, firstly on how a stack of two Kitaev chains, i.e the $2$-chain, is related to well-known SPT models. This is pictorially represented in Fig.~\ref{fig:equiv}, with details in the main text below. On the one hand we find a two-site unitary transforming the superconducting $2$-chain into the Su-Schrieffer-Heeger (SSH) model\cite{SSH}, a particle number preserving Hamiltonian with a complex fermionic edge mode protected by sublattice symmetry\cite{Ryu10}. This mapping arises naturally when using symmetry as a guide. Such a guiding principle even uncovers new facts in the case of known relationships, such as for the non-local transformation which maps the $2$-chain to the cluster model\cite{Suzuki71,Raussendorf01}, a spin-$\frac{1}{2}$ chain protected by a $\mathbb Z_2 \times \mathbb Z_2$ symmetry\cite{Son11}. Despite that mapping being well-known\cite{Keating2004,Son11,Smacchia11,Niu12,DeGottardi13,SON}, we uncover through it a new anti-unitary symmetry protecting the cluster model. This means that the cluster model and the Haldane phase are protected by the \emph{same} set of discrete symmetry groups, with a different microscopic action. Writing down a two-site unitary transforming one symmetry into the other, we map the cluster model to a spin chain whose ground state is the fixed point limit of the Affleck-Kennedy-Lieb-Tasaki (AKLT)\cite{AKLT} state. The AKLT model is a well-known perturbed spin-$1$ Heisenberg chain with an exactly known ground state.

Reconsidering the $\alpha$-chain also illuminates how the Haldane phase and the stack of four Kitaev chains, i.e. the $4$-chain, are two extremes of a single SPT model. The seminal work on the interacting classification of SPT phases showed that the $4$-chain has many algebraic similarities to the Haldane phase at the level of symmetries\cite{Fidkowski11}. We show that the interacting $4$-chain can locally be rewritten as a spinful Hubbard chain, protected by sublattice symmetry. In the Mott limit, it is a spin chain in the Haldane phase. Interestingly, in this limit the sublattice symmetry protecting the Hubbard chain is indistinguishable from spinful TRS. The latter is known to protect the Haldane phase, but \emph{only} in the Mott limit (even when combined with spin rotation symmetry)\cite{Anfuso2007}. Sublattice symmetry can be seen as a \emph{different} way of extending the same spin symmetry to charge degrees of freedom, in which case our construction shows that the Haldane phase remains stable despite charge fluctuations.

Other new physics arises from mapping the $\alpha$-chain to the \emph{generalized} spin-$\frac{1}{2}$ cluster models\cite{Suzuki71}. Such a relationship was observed before\cite{Keating2004,Smacchia11,Niu12,DeGottardi13,SON}, but we use it to uncover the SPT properties of these spin chains, for example leading to spin chains with \emph{both} symmetry breaking and SPT order. It is worth noting that despite these spin chains being mathematically equivalent to the $\alpha$-chains, their physics is distinct due to the non-locality of the mapping. This makes the set of cluster models useful in its own right (both for didactic and testing purposes), especially since one can add perturbations which break the equivalence to fermions yet leave the SPT properties intact.

In the second part of the paper, we use the $\alpha$-chain to explore the transitions between SPT phases (in the presence of interactions). We observe a direct relationship between the central charge describing the critical point and the topological properties characterizing both sides of the transition. We surmise that such a relationship holds for any topological phase transition described by a conformal field theory (CFT). In particular, if we interpolate between a trivial phase and an SPT phase with $d$-dimensional edge modes, we conjecture that the CFT describing the transition has a central charge $c \geq \log_2(d)$. We verify this conjecture for many known topological transitions, including all CFTs with central charge $c<1$ and certain classes of Wess-Zumino-Witten CFTs. The conjecture that $c\geq \log_2 d$ matches the idea that the gapless fields at the transition are the long-wavelength-fluctuations of a delocalized edge mode. Note that if this conjecture holds, it formalizes the intuition that the central charge measures the gapless degrees of freedom. Our conjecture can be seen as a far-reaching generalization of recent work\cite{Tsui17} which has shown that a transition between bosonic SPT phases satisfies $c \geq 1$.

An outline of the paper is as follows: in section \ref{scn:SPT} we present a brief review of one-dimensional SPTs, focusing on a physical perspective (with a systematic treatment given in Appendix \ref{app:sym}). Section \ref{scn:fSPT} concerns fermionic SPTs where we introduce the $\alpha$-chain and its symmetry fractionalization. In section \ref{sec:SSH} we illustrate how the $2$-chain is the SSH model in disguise, and in section \ref{sec:Hubbard} the interacting $4$-chain is adiabatically connected to the Haldane phase. We then turn to bosonic SPTs in section \ref{scn:sSPT} where the generalized cluster models emerge as the Jordan-Wigner transform of the $\alpha$-chain, pointing out how the physics has changed under this non-local mapping. This uncovers new non-trivial symmetries of the cluster model, which in section \ref{scn:cluster_AKLT} leads to identifying its ground state as the fixed point limit of the AKLT state. Section \ref{Kramers} shows how to generalize the Kramers-Wannier dualities to these generalized cluster models, shedding light on their symmetry breaking and SPT properties. Finally, in section \ref{scn:transition} we discuss the (interacting) phase transitions between these models, leading to the general conjecture which lower bounds the central charge at a critical point in terms of the edge modes of the gapped phases at both sides of the transition.

%%%%%%%%%%%%%%%%%%%%%%%%%%%%%%%%%%%%%%%%%%%%%%%%%%%%%%%%%%%%%%%%
\sectionheading{Symmetry protected topological phases}\label{scn:SPT}

Here we briefly review the classification of (interacting) SPT phases in 1D using physical pictures\cite{Fidkowski11,Turner-2010,Chen-2010,Schuch-2011}. First we present the general concept and then illustrate this in the case of the cluster model and Kitaev chain. For more details, we refer to Appendix \ref{app:sym} or the aforementioned references.

\textbf{Classification.} \hspace{5pt} Consider a gapped one-dimensional system of length $N$ invariant under a global symmetry group $G$. The classification scheme needs the symmetries to be well-defined even when having open boundaries\footnote{In case one is interested in spatial inversion symmetry, one has to replace the real-space picture by an entanglement-based approach.}, which for a unitary symmetry $U \in G$ is guaranteed if $U = \otimes_n U_n$ is a tensor product over sites or unit cells, referred to as an on-site symmetry. (The case of anti-unitaries, where complex conjugation is defined in some on-site/unit cell basis, is discussed in section \ref{scn:fSPT}.) If we assume $U$ is not spontaneously broken, then for periodic boundary conditions the ground state must be unique\footnote{This is clear if we assume our Hamiltonian is translation invariant. However, one does not strictly need translation invariance.} and hence invariant under $U$. However, if we have open boundary conditions, then the absence of spontaneous symmetry breaking in the bulk still allows for $U$ to act non-trivially near the edges. We write this as $U = U_L U_R$, which is valid in the ground state subspace. These effective operators $U_{L,R}$ are exponentially localized near the boundaries on a length-scale set by the correlation length. In the thermodynamic limit ($N \to \infty$) of a gapped phase, $U_L$ and $U_R$ thus have no overlap. Since the Hamiltonian is local, this means that $U_L$ and $U_R$ do not change the energy of a state in the ground state subspace. We refer to this as \emph{symmetry fractionalization}. The same holds for any other unbroken symmetry $V \in G$, so we can write $V = V_L V_R$. Any group relation between $U$ and $V$ then implies a relation between the edge symmetries. In particular, $\{U_L, V_L, \cdots\}$ then obey the same group relations as $\{U, V, \cdots\}$ possibly up to a phase factor. In the bosonic case, where $U_L$ and $U_R$ commute, both edges completely decouple and the physical symmetry is then \emph{projectively represented} on each edge. Such a projective representation has discrete labels that cannot change smoothly. Since any \emph{non-trivial} projective representation has a minimal dimension $>1$, it protects degenerate modes on the edge (which will be clear from the example of the cluster model). Fermions can have extra structures related to $U_L$ and $U_R$ not necessarily commuting, as will be clarified in the discussion of the Kitaev chain below.

\textbf{Bosonic example: the cluster model} \hspace{5pt} This is a spin chain with three-spin interactions:
\begin{equation}
H_C = -\sum_n X_{n-1} Z_n X_{n+1} \label{cluster}
\end{equation}
where we denote the Pauli spin operators $\sigma_{x,y,z}$ as $X,Y,Z$. Its earliest appearance in the literature is in Suzuki's work on quantum systems that are dual to two-dimensional classical dimer models\cite{Suzuki71} but it was reinvented by Raussendorf and Briegel in the context of measurement-based quantum computation\cite{Raussendorf01}. Keating and Mezzadri independently arrived at it as a spin chain dual to free fermions\cite{Keating2004} and Kopp and Chakravarty generated the model through a real space renormalization of the quantum Ising chain\cite{Kopp05}. The cluster model was discovered to be an SPT phase protected by $\mathbb Z_2\times \mathbb Z_2$ by Son et al.\cite{Son11}, and here we give a simplified treatment of this fact as found in Zeng et al.\cite{Wen_book}

If the total number of sites $N$ is even, $H_C$ is symmetric under the $\mathbb Z_2 \times \mathbb Z_2$ group generated by
\begin{equation}\label{def:P1P2}
\left\{
\begin{array}{ccl}
P_1 &= & Z_1 Z_3 Z_5 \cdots Z_{N-1} \\
P_2 &= & Z_2 Z_4 Z_6 \cdots Z_N
\end{array}
\right.
\end{equation}
Let us take open boundary conditions and analyze the edge modes. Note that the terms in Eq.~\eqref{cluster} square to one, such that the eigenvalues are $\pm 1$. Since all terms in $H$ commute, the ground state subspace will have $X_{n-1} Z_n X_{n+1} = 1$ for all $n$. Concatenating a list of these, we directly see that this implies $X_1 Z_2 Z_4 Z_6 \cdots Z_{N-2} X_{N-1} = 1$, or equivalently $P_2 = X_1 X_{N-1} Z_N$. So despite $P_2$ being a global operator, it turns out that in the ground state subspace it merely acts on the left by $P_2^L = X_1$ and on the right by $P_2^R = X_{N-1} Z_N$. Similarly, we obtain $P_1^L = Z_1 X_2$ and $P_1^R = X_N$.

We explicitly see that $P_1^L$ and $P_2^L$ are anticommuting symmetries, proving that the cluster model has degenerate edge modes. (Note that symmetry fractionalization generally only holds in the ground state subspace, whereas here $P_{1,2}^{L,R}$ commuting with the Hamiltonian imply so-called \emph{strong} zero modes\cite{Fendley16}.) Adding terms to Eq.\eqref{cluster} that respect the $\mathbb Z_2 \times \mathbb Z_2$ of $P_1$ and $P_2$ will alter the form of $P_1^L$ and $P_2^L$ but cannot change their mutual anticommutation: from $P_1 P_2 = P_2 P_1$ one can derive that $P_1^L P_2^L = e^{i\alpha} P_2^L P_1^L$, and from $P_1^2=1$ one can show that $ e^{i\alpha} = \pm 1$, indeed labeling the projective representations of $\mathbb Z_2 \times \mathbb Z_2$ (see Appendix \ref{app:sym} for details about symmetry fractionalization). Thus as long as the correlation length is finite, the edges have well-defined degeneracies. Hence the cluster model is an SPT phase protected by $\mathbb Z_2\times \mathbb Z_2$, however in section \ref{scn:sSPT} we will see how it relates to a longer-range Kitaev chain and how that uncovers new topological features and even a hidden $SO(3)$ symmetry in the ground state.

\textbf{Fermionic example: the Kitaev chain} \hspace{5pt} This is a one-dimensional model of superconducting fermions\cite{Kitaev-2001}:
\begin{equation}
H_K =\sum_n c_n^\dagger c_{n+1} + c_n^\dagger c_{n+1}^\dagger +\textrm{h.c.}
\label{Kchain} \end{equation}
Kitaev drew attention to this model in 2001 for the free Majorana modes on its edges. To see these, it is convenient to step away from the representation in terms of superconducting fermions and note that any fermionic mode can be decomposed into its real and imaginary part: $\gamma = c+c^\dagger$ and $\tilde \gamma = \frac{c-c^\dagger}{i}$. These Hermitian operators are Majorana modes, meaning they anti-commute and square to unity. One obtains the much simpler $H_K = i \sum_n \tilde \gamma_n \gamma_{n+1}$. Similar to the reasoning for the cluster model, the ground state subspace will have $\gamma_n \gamma_{n+1} = i$. This means that fermionic parity symmetry $P = \prod_i (1-2n_i) = \prod (i \tilde \gamma_n \gamma_n)$, which is a symmetry for any fermionic system, effectively acts as $P = i \gamma_1 \tilde \gamma_N$ for open boundaries. So here we see that $P = P_L P_R$ where $P_L$ and $P_R$ anticommute. So now we have a protected twofold degeneracy that is spread out over both edges. In other words there is a Majorana zero mode living on each edge, which can be said to be `$\sqrt{2}$-dimensional' -- \emph{by definition} this means that taking two such modes gives a $2$-dimensional Hilbert space. Because fermionic parity symmetry cannot be broken\footnote{By locality one would require that \unexpanded{$\langle c_i^\dagger c_j \rangle \xrightarrow{|i-j|\to \infty} \langle c_i^\dagger \rangle \langle c_j \rangle$}. Since the left-hand side is anti-symmetric and the right-hand symmetric, we require \unexpanded{$\langle c_i^\dagger \rangle = 0$}. Parity symmetry ensures this.}, this phase is stable under \emph{arbitrary} perturbations.

We see that if we only have $P$-symmetry, there are exactly two phases, characterized by $P_L P_R = \pm P_R P_L$. In the non-interacting classification this is referred to as the $\mathbb Z_2$ invariant of the D class\footnote{This requires a particle-hole symmetry, but note that \emph{any} fermionic Hamiltonian has a particle-hole symmetry as defined in the Bogoliubov-de Gennes Hamiltonian. In this paper we will not need this language.}. However, the Kitaev chain is also invariant under spinless time-reversal symmetry $T = K$, where $K$ is the complex conjugation that leaves $c$ and $c^\dagger$ invariant. If we enforce this symmetry, then in the absence of interactions we are put in the BDI class which is known to have $\mathbb Z$ distinct phases characterized by a certain topological invariant\cite{Ryu10}. However, in section \ref{sec:alpha} we review how in the interacting case there are only eight phases\cite{Fidkowski-2010}, labeled by $\mathbb Z_8$, where the topological invariants are constructed out of the symmetry fractionalization of $P$ and $T$\cite{Fidkowski11,Turner-2010,Bultinck17}. All the $\mathbb Z$ non-interacting phases (and the eight interacting ones) are generated by stacking single Kitaev chains. This uses the so-called group structure of SPT phases, which we explain now.

\textbf{Group structure of SPT phases.} \hspace{5pt} An important and elegant property of these phases is that the set of all SPT phases with respect to a given symmetry group $G$ \emph{itself} has a group structure. The addition of two SPT phases is defined by taking the physical stacking of both chains. This can be applied to both the non-interacting and interacting classification, but here we give the point of view relevant for the interacting classification, i.e. using symmetry fractionalization. For example, let $U$ be some unitary symmetry for a bosonic chain, then if we have $U = U^A_L U^A_R$ for the first chain and $U = U^B_L U^B_R$ for the second, then the combined system has $U_L = U^A_L U^B_L$. This new symmetry fractionalization will then be a realization of a possibly different SPT phase. It is not hard to convince oneself that this operation satisfies all the properties of a group. Mathematically, in the bosonic setting (where the edges fully decouple) this new group is called\cite{Chen-2010} the second group cohomology group of $G$ with coefficients in $U(1)$, denoted $H^2(G;U(1))$, although we do not use this language in this paper. For example, for $G = \mathbb Z_2 \times \mathbb Z_2$, the group of SPT phases is $\mathbb Z_2$: this means there is only one non-trivial phase --realized by the above cluster model-- which is its own inverse. Indeed, if one has a stack of two cluster models, then one can gap out the edge modes by the symmetry-preserving perturbation $V = \left(P_1^L\right)^A \left(P_1^L\right)^B + \left(P_2^L\right)^A \left(P_2^L\right)^B$.

\textbf{The subtlety of identifying phases.} \hspace{5pt} In the aforementioned, we did not distinguish the symmetry group from its representation. For example, the cluster model $H_C$ in Eq.~\eqref{cluster} has the abstract symmetry group $\mathbb Z_2 \times \mathbb Z_2$ which is \emph{represented} by $\{\mathbb I,P_1\} \times \{\mathbb I,P_2\}$. Indeed: in the classification scheme one usually \emph{fixes} the representation and only allows paths of gapped local Hamiltonians which are symmetric under that particular representation. Along such a path the symmetry fractionalization $\{\mathbb I,P_1^L\} \times \{\mathbb I,P_2^L\}$ remains well-defined. The downside of this definition is that \emph{any} other model with the same symmetry but in a different physical implementation is automatically in a distinct phase. Consider for example the spin-$\frac{1}{2}$ chain of alternating Heisenberg couplings $H = \sum_n \bm S_{2n} \cdot \bm S_{2n+1}$. The leftmost spin $\bm S_1$ is clearly decoupled, and this edge mode is in fact protected by any perturbation that preserves $\pi$-spin rotation since the edge transforms as a spin-$\frac{1}{2}$ whereas the bulk is a singlet. Thus it has the same symmetry group $\mathbb Z_2 \times \mathbb Z_2$ but now it is represented by $\{\mathbb I,R_x\} \times \{\mathbb I,R_y\} =\{\mathbb I,R_x,R_y,R_z\}$, sometimes referred to as the Haldane phase. According to the usual definition, these two models can \emph{not} be connected, despite both having the same properties with respect to $\mathbb Z_2 \times \mathbb Z_2$. To resolve this, we can introduce a broader notion of phase equivalence where one allows for paths of gapped local Hamiltonians where the symmetry group is preserved, but its on-site representation can vary smoothly. The symmetry fractionalization and consequent edge modes are then still protected quantities. In this way one can, for example, construct a path between the cluster model and the alternating Heisenberg chain where $P_1$ and $P_2$ smoothly transforms into $R_x$ and $R_y$. In section \ref{scn:cluster_AKLT} we naturally arrive at such a path purely from symmetry considerations, which moreover also preserves the other symmetries known to protect the Haldane phase. The condition that the symmetry remains on-site is crucial: if this is dropped, everything is trivial\cite{Schuch-2011}.

\sectionheading{Topological fermionic chains}\label{scn:fSPT}

\subsection{Stacking of Kitaev chains: the $\alpha$-chain} \label{sec:alpha}

\begin{figure}[h]
	\includegraphics[scale=.4]{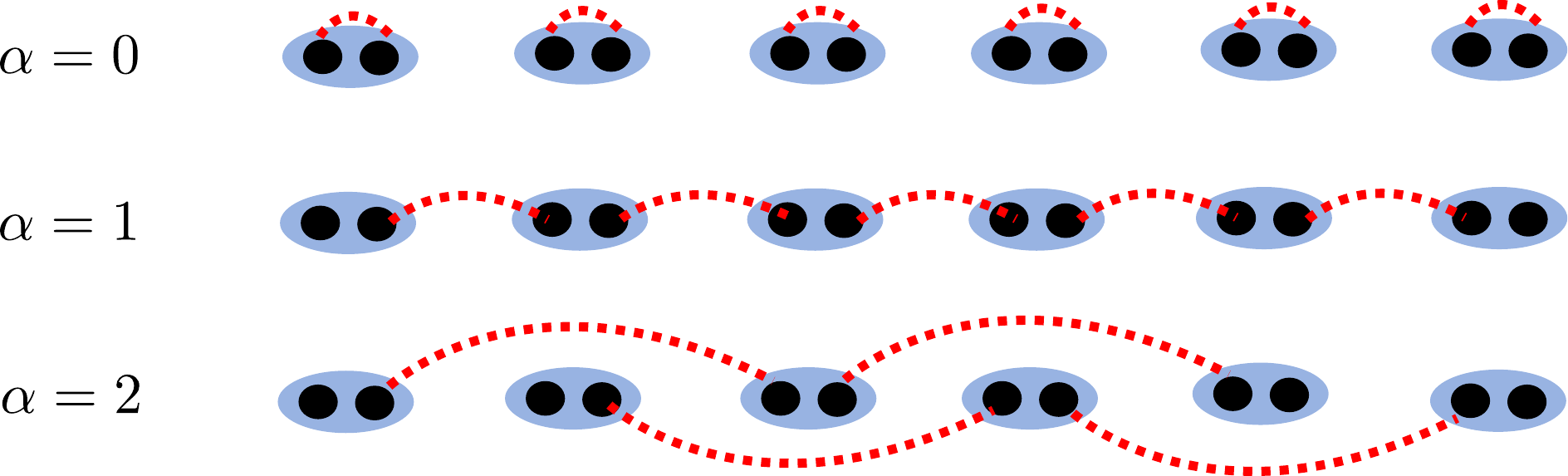}
	\caption{Schematically representing the $\alpha$-chain \eqref{alpha} for $\alpha=0,1,2$. The left black dot denotes the Majorana mode $\gamma$, the right one $\tilde \gamma$.} \label{fig:alpha}
\end{figure}

Here we reconsider the phases that arise by stacking multiple Kitaev chains. Instead of literally stacking them on top of one another, it is convenient to rewrite such stacks in a translation invariant manner. This makes it more natural, for example, to interpolate between a different number of chains without keeping systems artificially decoupled from one another. Fig.~\ref{fig:alpha} illustrates the idea: stacking $\alpha$ Kitaev chains on top of one another is equivalent to coupling the Majorana modes over a distance $\alpha \in \mathbb Z$. We call these $\alpha$-chains, with the Hamiltonian (where as before $\gamma = c+c^\dagger$ and $\tilde \gamma = \frac{c-c^\dagger}{i}$)
\begin{equation}
H_\alpha = i\sum_n \tilde \gamma_n \gamma_{n+\alpha} \label{alpha}
\end{equation}
Note that $H_1 = H_K$ and $H_0 = \sum_n 1-2c_n^\dagger c_n$, the trivial band insulator. This class of long-range Kitaev chains has been considered before in a non-interacting context\cite{DeGottardi13}. Their (interacting) SPT properties have been uncovered in the context of stacks of Kitaev chains\cite{Fidkowski11,Turner-2010}. Here we first revisit their SPT nature in an equivalent but slightly different language, before illuminating how the $\alpha$-chain maps to other models by local redefinitions. Non-local transformations to spin chains via a Jordan-Wigner transformation are discussed in section \ref{scn:sSPT}. We first discuss the topological nature of the $2$-chain before discussing the case for general $\alpha$.

\textbf{Symmetry fractionalization of the $2$-chain.} \hspace{5pt} For $\alpha = 2$, the left (right) has Majorana edge modes $\gamma_1,\gamma_2$ ($\tilde \gamma_N,\tilde \gamma_{N-1}$). These can be gapped out by the Hermitian perturbation $i\gamma_1\gamma_2$ ($i\tilde \gamma_N \tilde \gamma_{N-1}$), but this is not invariant under complex conjugation $T=K$. (Note that $T\gamma T = \gamma$ and $T \tilde \gamma T = - \tilde \gamma$ since we define $c$ and $c^\dagger$ to be invariant.) In fact, $T$ protects a Kramers pair on the right edge, and $PT$ on the left. To see this, let us define the fermionic edge modes
\begin{equation}
\left\{
\begin{array}{cl}
c_L &= \frac{1}{2}\left(\gamma_1 + i\gamma_2\right)\\
c_R &= \frac{1}{2}\left(\tilde\gamma_{N-1} + i\tilde\gamma_{N}\right)
\end{array} \right.
\end{equation}
It follows that $T c_L T = c_L^\dagger$ and $T c_R T = - c_R^\dagger$ (and oppositely for $PT$). From this one can derive that 
\begin{equation} \label{T2}
\left\{
\begin{array}{ccr}
T^2|0\rangle_L &= &|0\rangle_L\\
T^2|0\rangle_R &= &-|0\rangle_R
\end{array} \right.
\end{equation}
where we have defined $c_{L,R}|0\rangle_{L,R} = 0$. On first sight, this seems to contradict $T^2 = 1$, however performing the same calculation for \emph{both} edges gives $T^2 \left(|0\rangle_L \otimes |0\rangle_R\right) = |0\rangle_L \otimes |0\rangle_R$ (the extra minus sign coming from $c_L c_R = - c_R c_L$). These properties are summarized in row `$\alpha = 2$' of Table~\ref{table:alpha_symfrac}.

The fact that any fermionic Hamiltonian has parity symmetry $P$ begs the question whether the statement ``$T$ ($PT$) protects the right (left) edge'' has tangible consequences. To confirm it does, consider the Jordan-Wigner transformation which can map the $2$-chain to a spin model. This non-local transformation involves string operators which either start at the left or right edge. If the string originates from the right edge then local quantities near this edge will remain local in the new spin variables, hence $T$ protecting a Kramers pair implies a non-trivial spin chain protected by $T$. Oppositely, starting from the left edge should give a \emph{different} spin chain, now protected by $PT$. In section \ref{scn:sSPT} we see this is indeed the case.

As in the cases discussed above, this can be formulated in terms of symmetry fractionalization, which is slightly more subtle for anti-unitary symmetries. If we choose a basis for our low-energy space, then on these basis states $T$ acts as a unitary, with the fractionalization $T=U_L U_R$. Applying it twice, $1= T^2 = (TU_LT)(TU_RT)U_LU_R$. This means $TU_LTU_L = \pm 1$, with $TU_RTU_R$ having the same (opposite) sign if $U_{L,R}$ is bosonic (fermionic). These signs correspond to the square of $T$ on the edge modes, as in Eq.~\eqref{T2}. In particular, in the basis defined by $c_{L,R}$, we obtain $U_L = \gamma_2$ and $U_R = \tilde \gamma_{N-1}$ (the derivation and other details are discussed in Appendix \ref{app:sym}), such that indeed $TU_R T U_R = -1$, agreeing with Eq.~\eqref{T2}. The approach of previous works\cite{Fidkowski11,Turner-2010} was equivalent but different, opting for invariants which for $\alpha = 2$ would have $T$ square to $-1$ on the \emph{left} edge instead of the right. The above invariants strike us as more natural considering the physics of Eq.~\eqref{T2} and the ensuing discussion. Curiously, a recent approach\cite{Bultinck17} in terms of fermionic matrix product states does not have a spatial asymmetry in the fractionalization of $T$ (which suggest it might be implicitly describing a Jordan-Wigner transformed chain, cf. section \ref{scn:sSPT}).

\textbf{Symmetry fractionalization of the $\alpha$-chain.} \hspace{5pt} Stacks of Kitaev chains have played an important role in elucidating the classification of interacting SPT phases in one dimension\cite{Fidkowski11,Turner-2010} and it was realized that if we enforce $P$ and $T$ symmetry, there are eight possible phases. These correspond to $\alpha = 0,1,\cdots 7$ forming the group $\mathbb Z_8$ under SPT addition. In particular, Kitaev and Fidkowski\cite{Fidkowski-2010} have shown that a stack of eight non-interacting chains can be smoothly connected to a trivial phase if one allows for interactions, i.e. not just quadratic terms. Subsequently the eight remaining possibilities have been understood in terms of symmetry fractionalization, proving their stability against symmetry-preserving interactions. We summarize the result (derived in Appendix \ref{app:sym}) in Table~\ref{table:alpha_symfrac}, using the cyclic nature of $\mathbb Z_8$ to instead choose the representatives $\alpha = -3,-2,\cdots,4$ where the Hamiltonian \eqref{alpha} shows that negative $\alpha$ is the same as a left-right inverted $|\alpha|$-chain. For $-3 \leq \alpha \leq 3$, the low-energy subspace of one edge is too small to define interaction terms, hence they have the edge degeneracies we expect from the free fermion picture. For $\alpha = 4$ it was first discussed in Ref.~\onlinecite{Fidkowski-2010} how the perturbation $\gamma_1\gamma_2\gamma_3\gamma_4$ lifts the fourfold degeneracy on the left edge of the $4$-chain to a twofold degeneracy, which cannot be lifted further due to time-reversal symmetry.

\begin{table}[h]
	\begin{tabular}{c|c|c|c||c}
		$\alpha$ & $P$ & $T$ & $PT$ & total degeneracy \\ 
		\hline $-3$ & non-local fermion & left, (right) & [left], right & $8$\\
		$-2$ &  & fermion on left & right & $4$\\
		$-1$ & non-local fermion & (left) & [right] & $2$\\
		$0$ & & &  & $1$\\
		$1$ & non-local fermion & [right] & (left) & $2$\\
		$2$ & & right & left & $4$ \\
		$3$ & non-local fermion & [left], right & left, (right) & $8$\\
		$4$ & & left, right & left, right & $4$
	\end{tabular}
	\caption{The protected edge degeneracies for the $\alpha$-chains \eqref{alpha}. It also specifies where each symmetry protects a mode: `non-local fermion' means $P_L$ anti-commutes with $P_R$ (as discussed for the Kitaev chain in section \ref{scn:SPT}), whereas for example `$T$ left' means $T$ protects a Kramers pair on the left edge. Round and square brackets correspond to inequivalent choices of fractionalizing complex conjugation in the presence of a non-local mode, i.e. for a given choice one only has one of the two (details in Appendix~\ref{app:sym}). Different choices for distinct $\alpha$ still lead to invariants that distinguish the phases.}
	\label{table:alpha_symfrac}
\end{table}

\textbf{Left-right asymmetry.} \hspace{5pt} One peculiar feature of Table~\ref{table:alpha_symfrac} is the spatial asymmetry of the symmetry protection, which is possible due to the explicit inversion symmetry breaking of the model \eqref{alpha}: swapping left and right does not leave it invariant, but insteads changes the sign of $\alpha$. (We note that it is impossible in bosonic SPT phases for different edges to be protected by different symmetries, however it is possible for them to have different projective representations on each edge\cite{Morimoto14}.) For $\alpha = 4$, however, we see the same symmetries protect each edge, and indeed the fractional symmetries turn out to be bosonic. This means it cannot be represented in a free fermion system. In fact, as we discuss in section \ref{sec:Hubbard}, it can be seen as a spin SPT phase. Note that none of these eight phases can be connected by a path of gapped local Hamiltonians preserving $P$ and $T$. However $\alpha \leftrightarrow - \alpha$ are in the same phase according to the alternative notion discussed in section \ref{scn:SPT}, allowing paths which smoothly change the (on-site) representation of the anti-unitary symmetry from $T$ to $PT$.

\textbf{$\bm{O(|\alpha|)}$ symmetry of the $\alpha$-chain.} \hspace{5pt} Here we briefly discuss a symmetry which will be useful for what follows. As had first been observed in Ref.~\onlinecite{Fidkowski-2010}, a stack of $\alpha$ decoupled Kitaev chains has an $O(|\alpha|)$ symmetry. Conceptually this corresponds to rotating the chains into one another. On a mathematical level this is easy to see: if $\alpha > 0$ we define $\bm{\gamma}_n = (\gamma_{\alpha n}, \gamma_{\alpha n +1} \cdots, \gamma_{\alpha n + (\alpha -1)})$ and similarly $\bm{\tilde \gamma}_n$, since then $H_\alpha =i\sum_n \bm{\tilde \gamma}_n \cdot \bm \gamma_{n+1}$. The Hamiltonian is invariant under the linear action of $O(\alpha)$ on the vectors and this rotation preserves the Hermitian nature and canonical commutation relations $\{ \gamma_i,\gamma_j\} = 2\delta_{ij}$ and $\{ \gamma_i,\tilde \gamma_j\} = 0$.

\textbf{The $2$- and $4$-chain: SSH and Haldane.} \hspace{5pt} In the following two subsections we focus on the cases $\alpha = 2$ and $4$ respectively, uncovering their relationships to other known and new models. We first summarize some relevant observations of Fidkowski and Kitaev\cite{Fidkowski11}: firstly, as we have seen for $\alpha = 2$, each edge has a single complex fermionic zero mode. This is the same physics as the Su-Schrieffer-Heeger (SSH) model\cite{SSH}, whose Hamiltonian and properties we will soon discuss. Secondly, for $\alpha = 4$, the aforementioned $O(4)$ symmetry was realized to have an $SO(3)$ subgroup that acts projectively on the boundary. Combined with the non-trivial anti-unitary symmetry (cf. Table~\ref{table:alpha_symfrac}), the $4$-chain was henceforth labeled as being in the Haldane phase, a spin SPT phase with the same algebraic structure. We discuss these statements in more detail in sections \ref{sec:SSH} and \ref{sec:Hubbard}.

We show that these connections can be made surprisingly simple and concrete, which we summarize here before going into detail: in section \ref{sec:SSH} the $2$-chain in fact \emph{coincides} with the SSH model after a two-site change of basis. Moreover, this then implies the $4$-chain can be seen as a spinful SSH model. In section \ref{sec:Hubbard} we use this to rewrite the \emph{interacting} $4$-chain as a Hubbard model of spinful fermions where in the Mott limit the charge degrees of freedom are frozen out and the effective spin-$\frac{1}{2}$ model simply has alternating Heisenberg bonds (without any phase transition). Its ground state is the fixed point limit of the Affleck-Kennedy-Lieb-Tasaki (AKLT)\cite{AKLT} state, a canonical example of the Haldane phase. This leads to a much simplified constructive proof of the $8$-chain being adiabatically connected to a trivial chain.

It is interesting to keep track of the symmetries in the case of the $4$-chain, since this gives us insights into the stability of spin SPT phases when deviating away from the Mott limit (i.e. in the presence of charge fluctuations). In section \ref{sec:Hubbard} we rewrite the interacting $4$-chain as a so-called bipartite Hubbard model, which is known to have an $SO(4)$ symmetry\cite{Yang90}. This coincides with the $SO(4) \subset O(4)$ symmetry of the non-interacting $4$-chain mentioned above, surviving interactions. However, the $SO(3) \subset SO(4)$ which is realized projectively on its edge, does \emph{not} correspond to spin rotation symmetry in the language of the Hubbard model. Nevertheless, in the Mott limit this $SO(3)$ does become \emph{indistinguishable} from spin rotation symmetry (protecting the Haldane phase). This is a saving grace: it is known\cite{Anfuso2007} that the Haldane phase is not stable under charge fluctuations if one preserves spin rotation symmetry, however as a by-product of our construction we see the Haldane phase \emph{is} protected by this \emph{other} $SO(3)$ symmetry (or a $\mathbb Z_2 \times \mathbb Z_2$ subgroup thereof). Similarly, we will see the Hubbard chain is protected by an anti-unitary sublattice/particle-hole symmetry, in the Mott limit coinciding with the usual spinful time-reversal symmetry protecting the Haldane phase.

This underlines the fact that whether or not a spin SPT phase is stable under charge fluctuations (i.e. away from the Mott limit) depends on how the symmetry acts on the charge degrees of freedom. In particular, in the Mott limit, the Haldane phase can equivalently be said to be protected by spinful time-reversal \emph{or} fermionic sublattice symmetry --their action being indistinguishable. It is only \emph{away} from the Mott limit that the latter --not the former-- continues to protect the phase.

\subsection{The $2$-chain is the Su-Schrieffer-Heeger model} \label{sec:SSH}

We now relate the $2$-chain to the Su-Schrieffer-Heeger (SSH)\cite{SSH} model. The latter is a fermionic chain with a $U(1)$ particle conservation symmetry (its Hamiltonian given by Eq.~\eqref{hamiltonian:SSH}). On the other hand, the $2$-chain has superconducting terms, but as has been pointed out above it does have an $O(2)$ symmetry. Indeed if $\bm{\gamma}_n = (\gamma_{2n-1},\gamma_{2n})$ and $\bm{\tilde \gamma}_n = ( \tilde\gamma_{2n-1},\tilde \gamma_{2n})$, then $H_2=i\sum_n \bm{\tilde \gamma}_n \cdot \bm \gamma_{n+1}$. By relabeling some operators, we should be able to let the $SO(2) \subset O(2)$ act as a fermionic $U(1)$ phase symmetry. Note that this will have to involve mixing $\gamma$ and $\tilde \gamma$, since if $c\to e^{i\alpha} c$ and $\gamma = c+c^\dagger$ then $\gamma \to \gamma \cos \alpha - \tilde\gamma \sin\alpha $. Let us make this more precise for the interpolation between the $2$-chain and the trivial chain:
\begin{equation}
H_\textrm{SSH} = (1-\lambda)\sum_n  i\; \tilde \gamma_n \gamma_{n} + \lambda \sum_n  i\; \tilde \gamma_n \gamma_{n+2} \label{H}
\end{equation}
We then define $A$ and $B$ sublattices (i.e.: $A$ $B$ $A$ $B$ $A$ $B$ ...) and consider the new Majorana operators:
\begin{equation}\left\{ 
\begin{array}{lll}
\gamma_{A,n} &:=& \gamma_{2n} \\
\tilde \gamma_{A,n} &:=& \gamma_{2n-1} \\
\gamma_{B,n} &:=& \tilde \gamma_{2n-1} \\
\tilde \gamma_{B,n} &:= &-\tilde \gamma_{2n}
\end{array}\right.\label{def:SSH}\end{equation}
In terms of the complex fermionic operators consisting of these new Majorana operators (i.e. $c_{A,n} = \frac{1}{2} (\gamma_{A,n} + i\tilde \gamma_{A,n})$), we obtain
\begin{equation}
H_\textrm{SSH} = 2\sum_n \left[ (1-\lambda) \; c_{A,n}^\dagger c_{B,n}  + \lambda \; c_{A,n+1}^\dagger  c_{B,n} +\textrm{h.c.}  \right] \label{hamiltonian:SSH}
\end{equation}
\begin{figure}[h]
	\includegraphics[scale=.4]{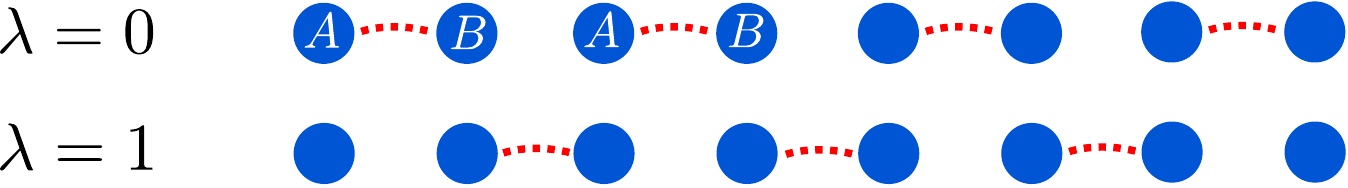}
	\caption{SSH model \eqref{hamiltonian:SSH} with fermionic edge modes for $\lambda=1$}\label{fig:SSH}
\end{figure}

This is exactly the Hamiltonian of the SSH model, schematically shown in Fig.~\ref{fig:SSH}. This undergoes a quantum phase transition at $\lambda = \frac{1}{2}$ and has protected edge modes for $\lambda > \frac{1}{2}$. What about the symmetries protecting it? Relabel the $2$-chain symmetries as $\mathcal C_B := T$ and $\mathcal C_A := PT$. From Eq.~\eqref{def:SSH} we see that the way they act on these new variables is as follows:
\begin{equation}
\mathcal C_B \; c_A \; \mathcal C_B = c_A^\dagger \qquad \textrm{and} \qquad \mathcal C_B \; c_B\; \mathcal C_B = -c_B^\dagger \label{sym}
\end{equation}
and similarly for $\mathcal C_A$. So our anti-unitary symmetries are particle-hole/sublattice symmetries. Despite $\mathcal C_B$ acting as a commuting anti-unitary symmetry on the Fock space Hamiltonian, one can check that it acts like an anti-commuting unitary on \emph{single-particle} Hamiltonians, i.e. this corresponds to the sublattice symmetry used in the non-interacting classification. Transposing our knowledge of the symmetry fractionalization of the $2$-chain, we know that for $\lambda =1$ the $\mathcal C$ symmetries fractionalize with $\mathcal C_A$ protecting the fermionic mode on the left, and $\mathcal C_B$ similarly on the right (and which on general grounds must be stable until the critical point at $\lambda = \frac{1}{2}$). This also tells us that the non-interacting label $\mathbb Z$ for the AIII class reduces to $\mathbb Z_4$ in the presence of interactions. It is worth noting that the $\alpha$-chains are stable under disorder whereas the SSH model is not (due to it requiring a sublattice symmetry), which is consistent with Eq.~\eqref{def:SSH} mixing neighboring sites.

\textbf{Identifying the two models.} \hspace{5pt} In effect the transformation \eqref{def:SSH} defines a local unitary $U$ that maps the $2$-chain to the SSH model. Since this unitary only acts \emph{within} the unit cells, we know that $\mathcal H$, defined by $U= e^{i \mathcal H}$, also only acts within the unit cells. Hence one can define the local unitary evolution $U(\lambda) = e^{i\lambda \mathcal H}$ which smoothly connects the models at the level of the Hamiltonian. It gradually deforms the representation of the anti-unitary symmetry from $T$ to $\mathcal C_B$, the crucial fact being that everywhere along this path the symmetry remains on-site (which for complex conjugation we take to mean that the basis it is defined in is on-site), which ensures that the symmetry fractionalization is everywhere well-defined. This is enough to say both models are in the same phase, as discussed in section \ref{scn:SPT}. The stronger statement that the unitary acts solely within the unit cells can be interpreted as the models not merely being in the same phase, but being virtually identical. To appreciate these facts, contrast it to the Kitaev chain mapping to the trivial chain under the local mapping $\gamma_n \to \gamma_{n-1}$, which one \emph{cannot} implement by a local unitary evolution. Or how the $2$-chain can be mapped to the trivial chain by a local unitary evolution, but such a path cannot keep the representations of the symmetries to be on-site.

\subsection{The $4$-chain as a Hubbard model and the AKLT chain}
\label{sec:Hubbard}

\begin{figure}[h!]
	\includegraphics[scale=.5]{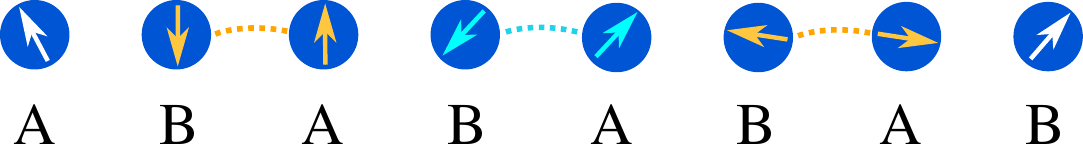}
	\caption{The bipartite Hubbard chain, Eq.~\eqref{Hubbard}, with single occupancy. For $U=0$ this is a double copy of the SSH state ($\lambda = 1$) in Fig.~\ref{fig:SSH}. For $U\to \infty$ this reduces to the AKLT fixed point state in Fig.~\ref{fig:AKLT}.} \label{fig:Hubbard}
\end{figure}

To gain insight into the interacting $4$-chain, we first show how it can be rewritten as a bipartite Hubbard model, smoothly connecting to a simple spin chain in its Mott limit. As we noted above, the $4$-chain does not have a non-interacting representation without accidental degeneracies: the perturbation $\gamma_1 \gamma_2 \gamma_3 \gamma_4$ lifts the fourfold degeneracy of the left edge into a twofold one, which according to Table~\ref{table:alpha_symfrac} cannot be lifted further if we preserve $P$ and $T$. So let us consider $H_\textrm{Hub} = \frac{1}{2} \left( H_4 + V + \tilde V \right)$ where
\begin{equation}
	 V = \frac{U}{2} \sum_{m=1}^{\frac{N}{4}} \gamma_{4m-3}\gamma_{4m-2}\gamma_{4m-1}\gamma_{4m}  \label{4chain}
\end{equation} 
and $\tilde V$ with $\gamma \leftrightarrow \tilde \gamma$.
The key idea is that we should be able to see the $4$-chain as a stack of two SSH chains, or alternatively as a single SSH model with an extra spin-$\frac{1}{2}$ degree of freedom. To make this more explicit, we first redefine $\gamma_{2n} = \gamma_{n,\downarrow}$ and $\gamma_{2n-1} = \gamma_{n,\uparrow}$ and then perform the transformation shown in Eq.~\eqref{def:SSH} for each spin sector. We summarize for clarity:
\begin{equation}\label{def:Hubbard}
\left\{ 
\begin{array}{lll}
c_{A,n,\uparrow} = \frac{1}{2} \left( \gamma_{4n-1} + i \gamma_{4n-3}\right) & & c_{A,n,\downarrow} = \frac{1}{2} \left( \gamma_{4n} + i \gamma_{4n-2}\right)\\
c_{B,n,\uparrow} = \frac{1}{2} \left( \tilde \gamma_{4n-3} - i \tilde \gamma_{4n-1}\right) & & c_{B,n,\downarrow} = \frac{1}{2} \left( \tilde \gamma_{4n-2} - i \tilde \gamma_{4n}\right)
\end{array}\right.
\end{equation}
In these new variables we obtain
\begin{align}\label{Hubbard}
H_\textrm{Hub} =&\sum_{n,\sigma} c^\dagger_{A,n+1,\sigma} c_{B,n,\sigma} + \textrm{h.c.} \\
& + U \sum_{\lambda,n} \left(n_{\lambda,n,\uparrow} - \frac{1}{2} \right)\left(n_{\lambda,n,\downarrow} - \frac{1}{2} \right) \nonumber
\end{align}
where $\lambda = A,B$ is the sublattice index. So we see the interacting $4$-chain is in fact a bipartite Hubbard chain, shown in Fig.~\ref{fig:Hubbard}. We note that this topological chain was investigated in Ref.~\onlinecite{Manmana12} using Green's functions. As long as $U\neq0$ the edges will prefer single occupancy, giving a twofold degeneracy on each edge. It can straightforwardly be proven that the gap of Eq.~\eqref{Hubbard} does not close as we increase $U$ (see e.g. the discussion by Anfuso and Rosch\cite{Anfuso2007}), in the Mott limit giving an antiferromagnetic spin chain $H_\textrm{Hub} \xrightarrow{\textrm{large }U} \frac{4}{U} \sum_n \bm{S}_{B,n} \bm \cdot \bm S_{A,n+1}$. Its ground state is simply a string of disjoint singlets with free spin-$\frac{1}{2}$ modes on the edges.

\textbf{Relation to the AKLT model.} \hspace{5pt} The AKLT model\cite{AKLT} is given by the spin-$1$ Hamiltonian $H = \sum_n \bm{S}_n \bm \cdot \bm S_{n+1} + \frac{1}{3} \left( \bm{S}_n \bm \cdot \bm S_{n+1} \right)^2$. This is known to be in the same phase as the spin-$1$ Heisenberg chain, but its ground state is exactly known. In fact it is the \emph{same} as the ground state of the above (large-$U$) bipartite Hubbard chain, with an additional spin-$1$ projector on every `AB' unit cell. The projection is in a sense immaterial: it leaves the entanglement spectrum \emph{between} the unit cells unchanged, moreover the projector naturally disappears under the renormalization group flow as defined in Ref.~\onlinecite{Verstraete05}. In this sense one can say that the ground state of the above Hubbard chain in the Mott limit is exactly the fixed point limit of the AKLT state. This simple ground state is naturally in the Haldane phase: while the bulk is invariant under $SO(3)$ and $T_\textrm{spin}$ (both of which are non-projective when applied to the unit cells), the edges transform as spin-$\frac{1}{2}$'s. The topological invariants of that projective representation define the celebrated Haldane phase. However, if we instead look at the relevant symmetries from the fermionic perspective, a different story emerges.

\textbf{Hubbard chain protected by sublattice symmetry.} \hspace{5pt} As a direct spin-off of section \ref{sec:SSH}, we know the Hubbard chain is an SPT phase protected by the anti-unitary sublattice/particle hole symmetry $\mathcal C_A$ defined in \eqref{sym}, which leaves the spin degree of freedom untouched. On first sight this seems unrelated to the symmetries of the Haldane phase, however in the Mott limit this reduces to $T_\textrm{spin} = e^{i\pi S^y} K$ which is known to protect the edge modes. Indeed: if $\bm S \coloneqq \frac{1}{2} c^\dagger_s \bm\sigma_{ss'} c_{s'}$ then by Eq.~\eqref{sym} we see that $\mathcal C_{A,B}$ are anti-unitaries that map $\bm S \to - \bm S$. (Moreover for any SPT phase protected by $T_\textrm{spin}$, globally  $T_\textrm{spin}^2=1$, even if it squares to $-1$ on-site.) Hence we can say that in the Mott limit we cannot distinguish between  $T_\textrm{spin}$ and $\mathcal C_A$ (or $\mathcal C_B$). However, away from the Mott limit, their difference is essential, as we discuss now.

\textbf{Fragility versus stability of spin SPT phases.} \hspace{5pt} In Ref.~\onlinecite{Anfuso2007} it was shown that one can adiabatically connect the Haldane phase to the trivial phase if one allows for paths  with fermionic degrees of freedom (i.e. away from the Mott limit). This is possible even if one preserves $T_\textrm{spin}$, which was interpreted as a sign of fragility of (bosonic) SPT phases with respect to charge fluctuations. However, here we see there is no fragility if we replace $T_\textrm{spin}$ by $\mathcal C_A$. Let us briefly repeat the reason why the Haldane phase is not stable against charge fluctuations\cite{Rosch12,Sanjay15} in the presence of $T_\textrm{spin}$. The reason it \emph{is} protected in the Mott limit, is because $T_\textrm{spin}^2 = 1$ in the bulk --since we have an even number of spin-$\frac{1}{2}$s-- from which one can deduce that on the edge it has to square to $\pm 1$, giving us a well-defined discrete invariant. But if every site no longer has exactly one fermion, we instead have $T_\textrm{spin}^2 = P$, where $P$ denotes the parity of the number of fermions, from which one can argue that its square on the edge can be smoothly deformed from $-1$ to $1$. It is then clear why $\mathcal C_A$ \emph{does} manage to protect the edge modes: it \emph{always} squares to the identity. Hence, there is no fragility with respect to this symmetry.

\textbf{Hubbard chain protected by $\bm{\mathbb Z_2\times \mathbb Z_2}$.} \hspace{5pt} Instead of time-reversal symmetry, the Haldane phase can also be protected by rotation symmetry: the global $\pi$-rotations $R_x = e^{i\pi S_x}$ and $R_y = e^{i\pi S_y}$ form a $\mathbb Z_2 \times \mathbb Z_2$ group that fractionalizes as spin-$\frac{1}{2}$ representations on the boundaries, i.e. $R_x^L R_y^L = - R_y^L R_x^L$. However, similarly to above\cite{Rosch12,Sanjay15} this does not protect the phase under charge fluctuations due to $R_x^2 = P$. As in the previous paragraph, one might wonder: although $R_x$ and $R_z$ do not protect the SPT phase, there might be a $\mathbb Z_2 \times \mathbb Z_2$ symmetry of the Hubbard chain which in the Mott limit becomes indistinguishable from the above spin rotation symmetry. Indeed, we introduce two unitary symmetries $\tilde R_x$ and $\tilde R_y$ which \emph{always} obey a $\mathbb Z_2 \times \mathbb Z_2$ group structure and in the Mott limit reduce to $R_x$ and $R_y$. This automatically proves they protect the Hubbard ladder for arbitrary interaction $U$. We define $\tilde R_x$ and $\tilde R_y$ as the unitary operators that square to one and act as:
\begin{equation}
\tilde R_x \; c_{\uparrow}\; \tilde R_x \; = \; c_{\downarrow}\;, \; \;
\tilde R_y \; \left\{ \begin{array}{c}
c_{A,\sigma}  \\
c_{B,\sigma}
\end{array} \right\} \; \tilde R_y = \left\{ \begin{array}{r}
c_{A,\sigma}^\dagger  \\
-c_{B,\sigma}^\dagger
\end{array} \right\} \label{def:Z2Z2}
\end{equation}
Then $\tilde R_x$ maps $\bm S \to (S_x,-S_y,-S_z)$ and $\tilde R_y$ maps $\bm S \to (-S_x,S_y,-S_z)$. Hence for large $U$ the actions of $\tilde R_x$ and $\tilde R_z$ are indistinguishable from $R_x$ and $R_y$. In conclusion, in the large $U$ limit we can identify the symmetries $R_x$, $R_y$ and $T_\textrm{spin}$ with $\tilde R_x$, $\tilde R_y$ and $\mathcal C_B$, but the latter set protects the Haldane phase even in the presence of charge fluctuations. (Moreover, note that $\mathcal C_B = \tilde R_y K$, extending $T_\textrm{spin} = R_y K$.)

It is known that the bipartite Hubbard model in fact has a much bigger on-site $SO(4)$ symmetry\cite{Yang90}. In terms of our original Majorana description in Eq.~\eqref{4chain}, if we define the vector $\bm{\gamma}_n = (\gamma_{4n-3},\gamma_{4n-2},\gamma_{4n-1},\gamma_{4n})$ and similarly $\bm{\tilde \gamma}_n$, then each element of $A \in SO(4)$ simply acts \textit{linearly} on this vector. Indeed, $H_4=i\sum_n \bm{\tilde \gamma}_n \cdot \bm \gamma_{n+1}$ is rotationally invariant and the interaction terms are of the form $\gamma_1 \gamma_2\gamma_3\gamma_4 = \epsilon_{i_1 i_2 i_3 i_4} \gamma_{i_1} \gamma_{i_2} \gamma_{i_3} \gamma_{i_4}$ such that $V \to \det(A) \; V$, thus the non-interacting $O(4)$ symmetry is broken down to $SO(4)$. The above $\mathbb Z_2 \times \mathbb Z_2$ symmetry group is a subgroup of this $SO(4)$: one can rewrite the action of Eq.~\eqref{def:Z2Z2} in terms of matrices which act linearly on the original Majorana variables, i.e.
\begin{equation}
\tilde R_x = \left(\begin{array}{cc} \sigma_x & 0 \\
0 & \sigma_x
\end{array}\right) \qquad \quad \tilde R_y = \left(\begin{array}{cc} -\mathbb I_2& 0 \\
0 & \mathbb I_2
\end{array}\right)
\end{equation}
In addition, we can write these in terms of generators of the Lie algebra $\mathfrak{so}(4)$, i.e. $\tilde R_{x,y} = \exp\left( i \pi \tilde S_{x,y} \right)$ where
\begin{equation}
\tilde S_x = \frac{i}{2}\left(\begin{array}{cc} 0 & - \mathbb I_2 + \sigma_x \\
\mathbb I_2 - \sigma_x & 0
\end{array}\right) 
\quad \;
\tilde S_y = \left(\begin{array}{cc} \sigma_y &0\\
0 & 0
\end{array}\right) 
\end{equation}
Note $\tilde R_z = \tilde R_x \tilde R_y = \exp\left( i \pi \tilde S_z \right)$ where $i\tilde S_z = [\tilde S_x,\tilde S_y]$. These operators satisfy the angular momentum algebra $[\tilde S_a,\tilde S_b] = i \varepsilon_{abc} \tilde S_c$, generating an $SO(3)$ subgroup of $SO(4)$. Thus there is the chain of symmetry groups $\mathbb Z_2\times \mathbb Z_2 \subset SO(3) \subset SO(4)$, each of which can be said to protect the edge modes.%(and note that each group only has \emph{one} non-trivial projective representation).
This agrees with the observation by Fidkowski and Kitaev that the $SO(4)$ symmetry of the interacting $4$-chain has an $SO(3)$ subgroup which transforms the edges under a spin-$\frac{1}{2}$ representation\cite{Fidkowski-2010}. In terms of the variables of the Hubbard chain \eqref{Hubbard}, in the Mott limit the above $SO(3)$ is indistinguishable from spin rotation acting on the unit cells.

\textbf{Connecting the $8$-chain to the trivial chain.} \hspace{5pt} We note that having connected the $4$-chain to the Haldane phase gives an alternative construction of an adiabatic path from a stack of eight Kitaev chains to the trival phase, which is considerably less technically involved than the original construction of Ref.~\onlinecite{Fidkowski-2010}. Interestingly, the path which we consider here in detail, was already sketched in Ref.~\onlinecite{Manmana12}. More precisely: one first tunes the $8$-chain to a stack of two decoupled spin chains with alternating (intra-chain) Heisenberg bonds. Now adiabatically turn off the intra-chain couplings and turn on the inter-chain Heisenberg couplings. This does not close the gap since it reduces to the four-spin problem $H = t \left( \bm S_1 \cdot \bm S_2 + \bm S_3 \cdot \bm S_4\right) + (1-t) \left( \bm S_1 \cdot \bm S_3 + \bm S_2 \cdot \bm S_4\right)$, whose distinct eigenvalues --there are maximally six due to $\frac{1}{2} \otimes\frac{1}{2} \otimes\frac{1}{2} \otimes\frac{1}{2} = 0 \oplus 0 \oplus 1 \oplus 1 \oplus 1 \oplus 2$-- can be obtained after some algebra, giving the gap $\sqrt{3\left(t-\frac{1}{2}\right)^2+\frac{1}{4}}$. The resulting phase is trivial: turning off the interactions leads us to a spinful version of Hamiltonian \eqref{hamiltonian:SSH} with $\lambda = 0$.

\textbf{Relation to previous work.} \hspace{5pt} As we have already mentioned at the end of section \ref{sec:alpha}, Fidkowski and Kitaev\cite{Fidkowski-2010} had observed that the algebraic properties of the interacting $4$-chain resemble those of the well-known Haldane phase. The new result here is that the path from the $4$-chain to such a spin chain is very simple and dictated by symmetries, directly leading to a close cousin of the AKLT model. This concrete path simultaneously raised and resolved the apparent paradox of the (in)stability of the Haldane phase with respect to charge fluctuations. We now point out the work of two other groups on the physics of the $4$-chain.

In 2012, Rosch\cite{Rosch12} showed how one can trivialize the $4$-chain (seen as superconducting spinless fermions) if one allows couplings to spin\emph{ful} fermions. More concretely, starting from a stack of a trivial spin chain and four Kitaev chains, a path to a completely trivial chain was constructed without breaking time-reversal symmetry (TRS). Indeed, if we only preserve TRS in the presence of charge fluctuations, we can first adiabatically transform our trivial spin chain to be in the Haldane phase. By the above, we now have a stack of \emph{two} Haldane chains, which can clearly be trivialized. Interestingly, that work defined variables very similar to the above \eqref{def:Hubbard} but did \emph{not} rewrite the Hamiltonian in terms of it. This is presumably due to a difference in philosophy: after defining the new variables, they were not seen as spinful fermions since the TRS of the original $4$-chain does not act as TRS on these variables. Our approach, however, is to consider \eqref{def:Hubbard} as defining genuine spinful fermions, and conclude that the Hubbard chain \eqref{Hubbard} is simply not protected by TRS but instead by the sublattice/particle-hole symmetry $\mathcal C_A$.

We also mention the field theoretic work of You et al.\cite{You2015a,You2015b}. They showed that starting from the $4$-chain, one can define spin operators out of these Majoranas whose effective continuum action upon integrating out the fermionic degrees of freedom is the \emph{same} non-linear sigma model that is known to describe the Haldane phase\cite{Haldane-1983}. Note that the work of Anfuso and Rosch\cite{Anfuso2007} has shown it can be subtle to draw conclusions about topological properties of a gapped phase after having integrated out other gapped degrees of freedom if these sectors were not completely decoupled to begin with. In the work of You et al. this decoupling is ensured by requiring the condensation of a particular $\mathbb Z_2$ gauge field. Without a physical mechanism to ensure this condensation (unlike the Hubbard chain \eqref{Hubbard} which ensures the gauge constraint $\gamma_1\gamma_2\gamma_3\gamma_4 = -1$ for large $U$), one cannot \emph{directly} transfer insights from the effective spin chain to the original fermionic one. It can however give very useful hints, and in Ref.~\onlinecite{You2015b} the knowledge of how to trivialize a stack of two Haldane chains was used to explicitly construct a path of the interacting $8$-chain to the trivial phase. Nevertheless, although this leads to a natural \emph{construction}, to actually \emph{confirm} the presence of a gap one still has to solve a rather complicated problem involving $16$ Majoranas, for which exact diagonalization (ED) was used. This is similar to the original path proposed by Kitaev and Fidkowski\cite{Fidkowski-2010}, where in addition to ED there was also a non-trivial analytic argument involving perturbation theory and the representation theory of $SO(8)$. Hence, to the best of our knowledge, having rewritten the $4$-chain as \eqref{Hubbard} has led to the simplest explicit path from the $8$-chain to the trivial chain, since it allows us to \emph{directly} use the spin chain results. It would be interesting to see if this approach can be helpful for the general program laid out in Ref.~\onlinecite{You2015b}, which elucidates the effect of interactions on fermionic SPT phases in higher dimensions by using known results for bosonic SPT phases.
 
\sectionheading{Topological spin chains}\label{scn:sSPT}

\subsection{The $\alpha$-chains map to generalized cluster models}

We now turn to spin SPT phases, focusing on spin chains which despite being \emph{mathematically} equivalent to the above fermionic $\alpha$-chains, are \emph{physically} quite distinct. To this purpose, recall that in one dimension, the non-local Jordan-Wigner transformation relates fermionic chains to spin-$\frac{1}{2}$ chains (with open boundary conditions) and vice versa:
\begin{equation}
\left\{ \begin{array}{l}
\gamma_n = Z_1 Z_2 \cdots Z_{n-1} X_n \\
\tilde \gamma_n = Z_1 Z_2 \cdots Z_{n-1} Y_n
\end{array} \right. \label{JW} \end{equation}
This transformation is compatible with the property that under complex conjugation ($T=K$) we have $T\gamma_n T = \gamma_n$ and $T \tilde\gamma_n T = -\tilde \gamma_n$. A priori it is not clear that such a non-local transformation preserves locality of the Hamiltonian. There is however a simple criterion: a fermionic Hamiltonian is local if and only if the corresponding spin Hamiltonian is local and commutes with spin-flip symmetry $P=\prod_n Z_n$. Let us now consider how Eq.~\eqref{JW} acts on our $\alpha$-chain \eqref{alpha}.

In the simplest case one can take the $0$-chain, which under Jordan-Wigner (JW) maps to a polarizing field $H = \sum_n Z_n$. More interesting is the well-known fact that the JW dual of the Kitaev chain $H_K$, i.e. the $1$-chain, is the Ising chain $H_I = -\sum_n X_nX_{n+1}$. This illustrates how, despite the JW transformation not changing the energy levels, the non-local mapping typically changes the physics: here it relates an SPT phase to a symmetry-broken phase. As a next step, consider the JW dual of the $2$-chain. Compared to the Kitaev chain, the Majorana operators are now one site further apart and hence one $Z$ of the JW string is not canceled, leading to the cluster model $H_C = -\sum_n X_{n-1}Z_nX_{n+1}$. This structure naturally extends to all $\alpha$-chains as shown in Table~\ref{table:JW}, where we see that the spatial inversion ($\alpha \leftrightarrow -\alpha$) on the fermionic side corresponds to $X\leftrightarrow Y$ on the spin side.

\begin{table}[h!]
	\begin{tabular}{c|c}
		Fermionic $\alpha$-chain \eqref{alpha} & Spin Hamiltonian after Jordan-Wigner \eqref{JW} \\
		\hline $\vdots$ & $\vdots$\\
		$\sum i\tilde \gamma_n\gamma_{n-2}$ & $-\sum Y_n Z_{n+1}Y_{n+2}$ \\
		$\sum i\tilde \gamma_n\gamma_{n-1}$ & $-\sum Y_n Y_{n+1}$ \\
		$\sum i\tilde \gamma_n\gamma_{n}\quad$ & $\sum Z_n$ \\
		$\sum i\tilde \gamma_n\gamma_{n+1}$ & $-\sum X_n X_{n+1}$ \\
		$\sum i\tilde \gamma_n\gamma_{n+2}$ & $-\sum X_n Z_{n+1}X_{n+2}$ \\
		$\vdots$ & $\vdots$\\
		$\sum i\tilde \gamma_n\gamma_{n+\alpha}$ & $-\sum X \underbrace{Z \quad \cdots \quad Z}_{\alpha-1}X$
	\end{tabular} 
	\caption{The $\alpha$-chain and its Jordan-Wigner transform}\label{table:JW}
\end{table}

These generalized cluster models first appeared in the literature in 1971 as the quantum chains dual to certain two-dimensional classical dimer models\cite{Suzuki71} (there referred to as generalized $XY$ models). In modern times they have resurfaced in studies of their phase transitions: first in context of exact results for their critical entanglement scaling\cite{Keating2004} and more recently concerning conjectures for their conformal field theories\cite{SON,Ohta16}. In section \ref{scn:transition} we return to the topic of these phase transitions from a different angle. On the other hand, it seems the physics of these \emph{gapped} spin chains has been left relatively unexplored. In particular, it is interesting to check how the physics of these spin models resembles or differs from the SPT structure of their fermionic counterparts. We will see these generalized cluster models exhibit rich physics despite their simplicity.

\textbf{The cluster model (`$\bm{XZX}$') and the $2$-chain.} \hspace{5pt} The special case of $\alpha =2$, the cluster model, is known\cite{Son11} to be in an SPT phase protected by the $\mathbb Z_2 \times \mathbb Z_2$ symmetry group generated by $P_1 = \prod_{n \textrm{ odd}} Z_n$ and $P_2 = \prod_{n \textrm{ even}} Z_n$. However, we now show that the mapping \eqref{JW} between the $2$-chain and the cluster model uncovers a hitherto-unknown symmetry which also protects the model. In section \ref{sec:alpha}, we saw that the right edge of the $2$-chain has a Kramers pair with respect to $T=K$, and the left edge with respect to $PT$. Since the Jordan-Wigner transformation \eqref{JW} has its string starting at the left edge, the leftmost region of the $2$-chain and the cluster model should have the \emph{same} local physics. We conclude that the cluster model has a Kramers pair on the left edge with respect to $PT = \left(\prod_n Z_n\right) K$. As discussed in section \ref{sec:alpha}, for a bosonic system an anti-unitary symmetry squares to the same sign on both edges. Hence, unlike the $2$-chain, $PT$ protects \emph{both} edges: the Jordan-Wigner transformation \eqref{JW}, which is highly non-local near the right edge, has changed the physics.

To see these statements within the spin language, first consider that section \ref{scn:SPT} implies $P = P_L P_R$ with $P_L = Y_1 X_2$ and $P_R = X_{N-1} Y_N$ (where we have used $P= P_1 P_2$). Secondly, in section \ref{Kramers} we show that the fractionalization of $T$ is trivial, i.e. $T = U_L U_R$ with $U_L = U_R = 1$ when acting on some local basis of edge states. The latter implies that when acting on this same basis, $PT = V_L V_R$ with $V_{L,R} = P_{L,R}$. Hence on the left $PT$ squares to $(PT) V_L (PT) V_L = - Y_1 X_2 Y_1 X_2 = -1$, and similarly on the right. This is summarized in row `$\alpha=2$' of Table~\eqref{table:spin}. The fact that $PT$ protects the cluster model explains why, for example, $H_C + \varepsilon \sum_n Y_n$ still has well-defined edge modes, as can be verified using perturbation theory or the numerical density matrix renormalization group (DMRG)\cite{White92} method. This new non-trivial symmetry can guide us to further insights -- which we discuss in section \ref{scn:cluster_AKLT}.

\begin{table}[h!]
	\begin{tabular}{c|c|c|c||c}
		$\alpha$ & $P$ & $T$ & $PT$ & total degeneracy \\ 
		\hline $-3$ & broken & Kramers pair: left, right & broken & $8$\\
		$-2$ &  & left, right & & $4$\\
		$-1$ & broken & broken & & $2$\\
		$0$ & & &  & $1$\\
		$1$ & broken & & broken & $2$\\
		$2$ & & & left,right & $4$ \\
		$3$ & broken & broken & left,right & $8$\\
		$4$ & & left, right & left, right & $4$
	\end{tabular}
	\caption{Symmetry breaking and fractionalization of the spin chains in Table~\ref{table:JW} with respect to $P = \prod Z_n$ and $T = K$. `Kramers pair on left' means the anti-unitary symmetry squares to $-1$ there.}
	\label{table:spin}
\end{table}	

\textbf{Symmetry fractionalization of the generalized cluster models.} \hspace{5pt} We now ask what the fractionalization is of these symmetries, $P$ and $T$, for any of these generalized cluster models. We repeat that this is different from the  fermionic results in Table~\ref{table:alpha_symfrac} since the non-local nature of the JW transformation mixes the edge with the bulk. Table~\ref{table:spin} was derived using the analytic methods introduced in section \ref{Kramers}, and numerically confirmed with DMRG\cite{White92} using the entanglement perspective discussed in Ref.~\onlinecite{Pollmann12b}. Note that the results are in line with what one would expect based on the Jordan-Wigner transformation: as discussed before, the Jordan-Wigner transformation whose string starts from the left end, should map the $2$-chain to an SPT protected by $PT$. Similarly, starting from the right end should map it to a spin model protected by $T$. This is the same as starting from the left end but taking the spatially inverted $2$-chain, i.e. the $(-2)$-chain, as confirmed by Table~\ref{table:spin}. Also note that Table~\ref{table:alpha_symfrac} says that --at least in a particular gauge-- the left end of the Kitaev chain ($\alpha = 1$) is protected by $PT$. This corresponds to the fact that the corresponding Ising chain spontaneously breaks $PT = \left( \prod_n Z_n \right)K$, whereas the dual of the $(-1)$-chain, $H = - \sum_n Y_n Y_{n+1}$, spontaneously breaks $T$.

The first symmetry of the resulting table is that like its fermionic dual, it only depends on $\alpha \mod 8$. The second symmetry is that swapping the $T$ and $PT$ columns is the same as changing the sign of $\alpha$: from Table~\ref{table:JW}, $\alpha \leftrightarrow -\alpha$ is equivalent to $X \leftrightarrow Y$, which is achieved by the anti-unitary operator $\mathcal O = e^{i\frac{\pi}{4} \sum_n Z_n}K$, and indeed $\mathcal O \; T \; \mathcal O = PT$.

\textbf{Symmetry breaking and/or SPT order} \hspace{5pt} Before discussing generalized cluster models for specific $\alpha$, let us observe their overall symmetry breaking and SPT properties. Every odd $\alpha$ has $\mathbb Z_2$ symmetry breaking. This is to be expected: the degeneracy ($=2^{\alpha}$) is then not a multiple of $4$, meaning we cannot associate it to bosonic modes on each edge. Hence there must be a degeneracy even with periodic boundaries, which for gapped phases in one dimension is always due to spontaneous symmetry breaking. In section \ref{Kramers} we show a general argument for the absence or presence of symmetry breaking that is purely self-contained in the spin language.

On the other hand, even $\alpha$ give rise to (purely) SPT phases. The four inequivalent even-$\alpha$ phases have a $\mathbb Z_2\times \mathbb Z_2$ structure:
%\footnote{One cannot give an algebraic structure to all eight spin phases, since stacks of symmetry breaking phases do not have well-defined order.}:
each is its own inverse, and stacking any two non-trivial chains generates the third. This is to be contrasted with the $\mathbb Z_8$ of the eight fermionic SPT phases. The non-local JW transformation does not commute with the procedure of stacking, in the sense that, for example, a stack of two cluster models does not correspond to a stack of two $2$-chains under JW.

The symmetries of Table~\ref{table:spin} imply that the only new phases (at least with respect to these symmetries) are $\alpha = 3,4$, since the negative $\alpha$ are related to positive $\alpha$ by a symmetry transformation. In fact the models related by $\alpha \leftrightarrow - \alpha$ are in the same phase if we allow for paths of gapped local Hamiltonians that smoothly change the on-site representation of the symmetries, transforming $T$ into $PT$ (where, again, by `on-site anti-unitary' we mean that the basis for complex conjugation is on-site). Hence, before turning to the cluster model in more detail in section \ref{scn:cluster_AKLT}, we discuss the physics of $\alpha = 3,4$.

\textbf{The `$\bm{XZZX}$' cluster model.} \hspace{5pt} Interestingly, $\alpha = 3$ has both symmetry breaking \emph{and} SPT order. In particular we find that the symmetry breaking order parameter\footnote{Indeed: a state with $XYX = \pm 1$ will satisfy $XZZX = 1$ since $\left( X_n Y_{n+1} X_{n+2} \right) \left( X_{n+1} Y_{n+2} X_{n+3} \right) = X_nZ_{n+1}Z_{n+2}X_{n+3}$.} is a cluster-type term, $X_{n-1} Y_n X_{n+1}$, such that a symmetry-broken sector has the effective Hamiltonian $H_\pm=\pm \sum_n X_{n-1}Y_nX_{n+1}$. This still has $PT$ as a symmetry and it turns out that its symmetry fractionalization is the same as for $\alpha = 2$. More generally: for odd $0<\alpha<4$, the $\alpha$-chain spontaneously breaks into a ground state sector which is in the same phase as the $(\alpha - 1)$-chain with respect to the unbroken symmetry (and similarly for negative $\alpha$). A particular manifestation is that the symmetry-broken ground state of the Ising chain is trivial.

\textbf{The `$\bm{XZZZX}$' cluster model.} \hspace{5pt} The case $\alpha = 4$ is again purely an SPT phase (and similar to the fermionic $4$-chain one needs extra terms to lift accidental degeneracies: the Jordan-Wigner transform of Eqn.~\eqref{4chain} gives terms of the form $X_nY_{n+1}X_{n+2}Y_{n+3} + (X\leftrightarrow Y)$). If one compares the symmetry fractionalization tables of the fermionic $\alpha$-chain (Table~\ref{table:alpha_symfrac}) and the generalized cluster models (Table~\ref{table:spin}), the only non-trivial line that coincides is exactly $\alpha = 4$. Hence one might be tempted to conclude these two are in the same phase. This is in fact not true, the fundamental reason being that the `$P$' in the fermionic case is fermionic parity symmetry, which is intrinsic to the Hilbert space, whereas the `$P$' in the spin models is spin-flip symmetry which one can explicitly break. More concretely: there can be no path of gapped local Hamiltonians connecting the fermionic $\alpha =4$ to the bosonic $\alpha = 4$, even if we allow the on-site representation of the relevant symmetries to smoothly change. The difference becomes even more striking: in the following section, we show how the cluster model is in fact in the Haldane phase with all its discrete symmetries. Combining this with section \ref{sec:Hubbard}, we know there is a path connecting the fermionic $4$-chain to the cluster model ($\alpha =2$), which then proves there cannot be a path to the generalized cluster model with $\alpha = 4$.

\subsection{The cluster state is the AKLT fixed point limit} \label{scn:cluster_AKLT}

The previous section showed that there are two sets of symmetries protecting the cluster model $H_C = -\sum_n X_{n-1}Z_nX_{n+1}$: firstly a pair of commuting unitary symmetries squaring to one ($P_1$ and $P_2$), and secondly an anti-unitary symmetry that squares to one ($PT$). For the SPT phase to survive, one needs to only preserve one of these sets. There is another well-known bosonic SPT phase with the same algebraic properties: the Haldane phase. As encountered in section \ref{scn:fSPT}, it is an SPT phase protected by either the group of $\pi$-rotations which \emph{in the bulk} square to one (generated by $R_x = e^{i\pi S_x}$ and $R_y = e^{i\pi S_y}$) or by the time-reversal symmetry that squares to one ($T_\textrm{spin} = R_y K$).

\begin{figure}[h]%[tb]
	\includegraphics[scale=.5]{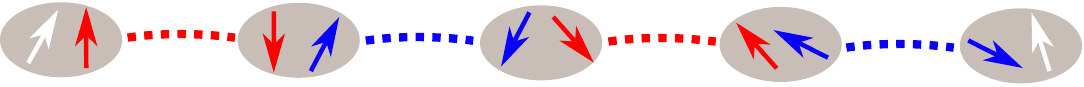}
	\caption{The AKLT state with dashed lines denoting spin singlets. The AKLT state has spin-$1$ projectors on the gray ovals\cite{AKLT}, disappearing in the fixed point limit\cite{Verstraete05}.}
	\label{fig:AKLT}
\end{figure}

This similarity is in fact not accidental: the cluster state is actually in the Haldane phase! For this to be a meaningful statement, we first need to perform a change of basis so that the symmetry operators map to each other:
\begin{equation}
P_1 \to R_x \qquad P_2 \to R_y \qquad PT \to T_\textrm{spin} \label{change_of_basis}
\end{equation}
Note that this is possible because the operators share the same group properties. It turns out that after this change of basis, the spin cluster ground state is actually mapped exactly to the fixed point limit of the AKLT state encountered before, sketched in Fig.~\ref{fig:AKLT}: each oval denotes a unit cell such that it has a linear representation of rotation and time-reversal symmetry. The dashed lines denote spin singlets on the bonds, with unconstrained spin-$\frac{1}{2}$'s on each edge, protected by the projective representations of the bulk symmetries.

\textbf{Identifying symmetries.} \hspace{5pt} More exactly, let us start with our spin-$\frac{1}{2}$ cluster Hamiltonian $H_C=-\sum_n X_{n-1}Z_nX_{n+1}$. Note that although this is translation invariant, the symmetry $P_1$ is not, so if we want a new basis where this symmetry acts as $R_x$, then we need to artificially work with unit cells of two spins. We now define a unitary operator $U$ which is a tensor product over these unit cells, acting in each cell as follows:
\begin{equation}
\begin{array}{ccr}
|\uparrow \uparrow \rangle & \qquad \xrightarrow{U} \qquad & |s \rangle \coloneqq \frac{1}{\sqrt{2}} (|\uparrow \downarrow \rangle - |\downarrow \uparrow \rangle ) \\
|\uparrow \downarrow \rangle & \xrightarrow{U} & |x \rangle \coloneqq \frac{1}{\sqrt{2}} (|\uparrow \uparrow \rangle - |\downarrow \downarrow \rangle ) \\
|\downarrow \uparrow \rangle & \xrightarrow{U} & -|y \rangle \coloneqq \frac{i}{\sqrt{2}} ( |\uparrow \uparrow \rangle + |\downarrow \downarrow \rangle) \\
|\downarrow \downarrow \rangle & \xrightarrow{U} & i|z \rangle \coloneqq \frac{i}{\sqrt{2}} ( |\uparrow \downarrow \rangle + |\downarrow \uparrow \rangle )  
\end{array} \label{basis}
\end{equation}
The labels $|s,x,y,z\rangle$ which we define on the right-hand side of Eq.~\eqref{basis} imply their symmetry properties, e.g. $|y\rangle$ goes to minus itself under $R_x$ or $T_\textrm{spin}$, but is invariant under $R_y$. Note that the unitary $U$ is naturally determined by the symmetry considerations of \eqref{change_of_basis}: if we, for example, apply $P_1$ on the left-hand side of Eq.~\eqref{basis}, then this is equivalent to applying $R_x$ on the right-hand side. More concretely, its defining characteristics are $U P_1 U^\dagger = R_x$, $U P_2 U^\dagger = R_y$ and $U (PT) U^\dagger = T_\textrm{spin}$, accomplishing \eqref{change_of_basis}. Moreover, note that this can be done smoothly, similar to as we discussed in section \ref{sec:SSH}.

\textbf{The resulting Hamiltonian.} \hspace{5pt} Having used symmetries to obtain the relevant change of basis, we can now see how it affects the cluster model. Curiously, the unitary has the effect of factorizing the Hamiltonian: e.g.\footnote{More precisely, $X_{2n}$ becomes $X_{2n}$ and $Z_{2n-1}X_{2n}$ becomes $-X_{2n-1}$. Similarly $X_{2n-1} \Rightarrow Y_{2n-1}$ and $X_{2n-1}Z_{2n} \Rightarrow -Y_{2n}$.} $-X_1 Z_2 X_3$ becomes $Y_2Y_3$, and $-X_2Z_3X_4$ becomes $X_2X_3$. Thus the Hamiltonian in this basis is a sum of \emph{disjoint} operators, which moreover turn out to be projectors:
\begin{align} \label{alternating}
U \; H_C \; U^\dagger &= \sum_n (X_{2n}X_{2n+1}+Y_{2n}Y_{2n+1}) \\
 &=  -\sum_\textrm{bond} (|s\rangle \langle s| -|z\rangle \langle z|) \notag
\end{align}
The ground state of this is the state with a singlet $|s\rangle$ on each \textit{bond} connecting the unit cells as in Fig.~\ref{fig:AKLT}. As mentioned in section \ref{sec:Hubbard}, it is obtained from the original AKLT state by a block-spin RG flow which does not change the bipartition entanglement spectrum\cite{Verstraete05}. An alternative way of checking that the cluster state and the fixed point limit of the AKLT state are the same is by comparing their matrix product state description\footnote{The MPS matrices for either state are $\mathbb I,X,iY,Z$. For the cluster state this is in the basis \unexpanded{$|\uparrow \uparrow \rangle, |\uparrow \downarrow\rangle, |\downarrow \downarrow \rangle, |\downarrow \uparrow \rangle$}. For the fixed point limit of the AKLT state this is in the \unexpanded{$|s\rangle,|x\rangle,i|y\rangle,|z\rangle$} basis.}. As an aside, note that the cluster state is translation invariant, but its symmetries have a two-site unit cell. The change of basis swaps this: the AKLT state (Fig.~\ref{fig:AKLT}) has a two-site unit cell, but its symmetries are on-site.

\textbf{Consequences.} \hspace{5pt} This mapping can teach us a few things: for example the Haldane phase is also known to be protected by link inversion symmetry, which is lattice inversion about the center of a bond. So we can conclude that the cluster state is similarly protected by such a symmetry\footnote{It is however a bit unnatural: one inverts the lattice of unit cells but not the unit cells themselves; moreover there is a $-1$ factor for every unit cell in state \unexpanded{$|\uparrow\uparrow\rangle$}. Example: \unexpanded{$|\downarrow\uparrow,\uparrow \uparrow \rangle \Rightarrow - | \uparrow \uparrow , \downarrow\uparrow \rangle$}. Usual lattice inversion does not protect the SPT phase.}. Moreover, it is known that the AKLT state is symmetric under continuous spin rotation. The fact that the cluster ground state must \emph{also} have an $SO(3)$ symmetry is a priori surprising, given its definition. Similarly this implies the $2$-chain and the SSH model ($\lambda = 1$), whose $O(2)$ symmetry we already discussed in section \ref{scn:fSPT}, has a ground state with $SO(3)$ symmetry. Note that this is completely unrelated to the symmetries we discussed in section \ref{scn:fSPT} having to do with rotating the Kitaev chais into one another: that concerned symmetries of the Hamiltonian, whereas this $SO(3)$ is an emergent symmetry in the ground state. (It is worth pointing out that one can adiabatically turn on the $Z_{2n} Z_{2n+1}$ component in \eqref{alternating} without affecting the ground state, until one reaches an $SO(3)$-symmetric Hamiltonian: the alternating Heisenberg chain we have encountered before in section \ref{scn:SPT} and \ref{sec:Hubbard}.) In the other direction, the cluster state has been mainly investigated in the context of its power for measurement-based quantum computation. It was only later that it was realized that the AKLT state\cite{Gross07} and more generally the Haldane phase\cite{Else12,Raussendorf12} offer a similar resource. Our mapping makes this more direct, and illustrates how by identifying symmetries one can construct natural maps that relate seemingly different models. Note that both the cluster state and the AKLT state have been generalized to 2D, both of particular interest to measurement-based quantum computing, and it would be of interest to see to what extent this kind of symmetry-guided mapping can generalize.

%\textbf{`$\bm{XZX}$' is `AKLT', is `SSH', is the $2$-chain}
\textbf{Identifying `$\bm{XZX}$', `AKLT', `SSH' and the $2$-chain} \hspace{5pt} In Fig.~\ref{fig:equiv} we summarize a few of the mappings related to the $2$-chain. In particular we complete the circle by noting that if we use the JW transformation to map back our spin model in the new basis to fermions, we obtain the SSH model. Let us take this step by step: starting with the linear interpolation between the trivial chain and cluster model,
\begin{equation}
H = (1-\lambda) \; \sum_n  Z_n  \; - \lambda \; \sum_n X_{n-1}Z_nX_{n+1}
\end{equation}
then under Eq.~\eqref{basis} this maps to the alternating spin-$\frac{1}{2}$ $XY$-chain (which moreover continuously connects to the alternating Heisenberg chain):
\begin{align} \label{XY}
UHU^\dagger = (1-\lambda) \sum_{n\textrm{ odd}} &(X_nX_{n+1}+Y_nY_{n+1}) \\
+ \; \lambda  \sum_{n\textrm{ even}} &(X_nX_{n+1}+Y_nY_{n+1}) \notag
\end{align}
Note that with respect to the unit cells which group together $(2m-1,2m)$, this is trivial for $\lambda < \frac{1}{2}$ and in the Haldane phase for $\lambda > \frac{1}{2}$. After the usual Jordan-Wigner map \eqref{JW}, Eq.~\eqref{XY} coincides with the SSH model as shown in Eq.~\eqref{hamiltonian:SSH}. Remember that the SSH model is protected by the sublattice/particle-hole symmetries $\mathcal C_{A,B}$ as defined in \eqref{sym}. One can check that $\mathcal C_A$ on the fermionic side (which protects the left edge) maps to $T_\textrm{spin}$ on the spin side (which protects both edges), and similarly $\mathcal C_B$ (which protects the right edge) maps to $P T_\textrm{spin}$ (which protects nothing). Again we see that the JW transformation changes the physics.

Note that Fig.~\ref{fig:equiv} does not contain the connection between the interacting $4$-chain and the AKLT state as discussed in section \ref{sec:Hubbard}, which implies that the fermionic $4$-chain can be adiabatically connected to the cluster model.

\textbf{`$\bm{XZZZX}$' is not in the Haldane phase} \hspace{5pt} Similarly one can subject the generalized cluster model with $\alpha = 4$ (which is also symmetric under $P_1$, $P_2$ and $T$) to the same mapping \eqref{basis}. We then obtain 
\begin{equation}
H = \sum_{n\textrm{ even}} \left( X_n Z_{n+1} Z_{n+2} X_{n+3} + 
Y_n Z_{n+1} Z_{n+2} Y_{n+3} \right)
\end{equation}This is now a spin chain with the \emph{same} discrete symmetries as the Haldane phase, i.e. $R_x, R_y$ and $T_\textrm{spin}$, yet it is in a \emph{different} symmetry class. It is protected by $T_\textrm{spin}$ --like the Haldane phase-- but \emph{also} by $R_{x,y,z} T_\textrm{spin}$ --unlike the Haldane phase. Moroever, it is \emph{not} protected by just the $\pi$-spin rotations. In particular one can derive $R_x^L = X_1Z_2Y_3$ and $R_y^L = Y_1Z_2X_3$, which clearly commute.
%\footnote{Alternatively, taking the basis $H = - \sum XZZZX$, one can show that $P_1^L = Y_1 Z_2 X_3$ and $P_2^L = Y_2 Z_3 X_4$.}
This shows it is very different from the fermionic $4$-chain, despite both on first sight sharing a similar symmetry fractionalization in Tables~\ref{table:alpha_symfrac} and \ref{table:spin}. This illustrates the physical subtleties of the Jordan-Wigner transform.

\subsection{Kramers-Wannier dualities for the generalized cluster models} \label{Kramers}

The generalized cluster models are all exactly soluble in terms of fermions, however often it can be cumbersome to extract the relevant information in the spin language. Here we present a way of extracting the physics we have discussed so far --directly in the spin language. Many properties simply drop out, such as the occurrence of spontaneous symmetry breaking (only) for $\alpha$ odd and the symmetry fractionalization of the topological phases. Concretely, we show how any of the generalized cluster models can be mapped to a trivial spin chain using a type of Kramers-Wannier transformation. The original transformation\cite{KW1941} is a duality of the quantum Ising chain which relates the symmetry-broken phase to the trivial phase and vice versa. Here we generalize this mapping, which in particular will show that for periodic boundary conditions the ground state is unique for $\alpha$ even and twofold degenerate for $\alpha$ odd, implying symmetry breaking. Note that before repeating the original mapping, we first treat the case where $\alpha$ is even since it is in fact simpler.

\textbf{$\bm \alpha$ even} \hspace{5pt} For clarity we take the cluster model (i.e. $\alpha = 2$) but the argument extends. Define the new spin operators $\tilde X_n = X_n$ and $\tilde Z_n = X_{n-1} Z_n X_{n+1}$. These indeed obey the Pauli algebra. Then $H_C = - \sum_n X_{n-1}Z_nX_{n+1} = - \sum_{n=1}^N \tilde Z_n$. Clearly the ground state is unique! Note that for open boundary conditions, $\tilde Z_1$ and $\tilde Z_N$ would not appear in the Hamiltonian, giving the correct edge degeneracies. In fact this allows for a slightly different derivation of the symmetry fractionalization, e.g. $P_1 = \prod_\textrm{odd} Z = \prod_\textrm{odd} \tilde Z = \tilde Z_1 = X_N Z_1 X_2$, which we already knew. However, it also allows to calculate other fractionalizations such as that of $T = K$: because in this case the mapping preserves complex conjugation and the ground state subspace condition $\tilde Z_{2 \leq n \leq N-1} = 1$ is also real, one easily obtains that $T = K'$, where $K'$ is complex conjugation in the low energy subspace. I.e. $T$ is trivial for the cluster state.

\textbf{$\bm \alpha$ odd} \hspace{5pt} Inspired by the above, one might similarly define $\tilde Z_n = X_n X_{n+1}$ for $H=-\sum_n X_nX_{n+1}$, but then there is no choice of $\tilde X_n$ that satisfies the Pauli algebra. However, if we redefine $\tilde Z_N = X_N$ for the last site, then choosing the domain-wall operators $\tilde X_n = Z_1 Z_2 \cdots Z_n$ gives the correct algebra. For periodic boundary conditions we obtain
\begin{equation}
H_I = - \sum_{n=1}^N X_nX_{n+1} = - \sum_{n=1}^{N-1} \tilde Z_n - \prod_{n=1}^{N-1} \tilde Z_n \label{KW}
\end{equation}
Now the ground state is clearly twofold degenerate. This constructions works for all odd $\alpha$ by extending $\tilde Z_n = X_n Z \cdots Z X_{n+\alpha}$ and $\tilde Z_N = X_N Z_1 \cdots Z_{\alpha -1}$, which indeed defines a consistent Pauli algebra with $\tilde X_n = \prod_{k=1}^n X_k Z\cdots Z X_{k+\alpha-1}$. The Hamiltonian is again of the form of Eq.~\eqref{KW} with a twofold degeneracy. Note that for open boundary conditions, the harmless product term drops away and $\alpha-1$ terms disappear from the sum, giving a $2^\alpha$-fold degeneracy for these pure models.

%%%%%%%%%%%%%%%%%%%%%%%%%%%%%%%%%%%%%%%%%%%%%%%%%%%%%%%%%%%%%%%%
\sectionheading{Transitions between topological phases}\label{scn:transition}

\subsection{Goal: predicting properties of transitions between topological phases}
We now investigate the transitions between one-dimensional SPT phases. We are guided by the question ``\emph{given two topological phases, can one predict the universal properties of the critical point between them?}''. Here we are interested in the situation where there is a direct transition, i.e. no intermediate phase, with the critical point being described by a conformal field theory (CFT). The examples we will discuss show that this is in fact a common situation, although we do not enter the discussion of whether this is more (or less) generic than first order transitions or intermediate phases.

By first using the above $\alpha$-chain model as a representative testing ground, we arrive at a general conjecture which formulates a partial (affirmative) answer to the above question, which we then check in other cases. The conjecture relates the central charge of the CFT --which counts the gapless degrees of freedom at the critical point-- to the topologically protected edge zero modes in the neighboring gapped phases.

\textbf{Central charge} \hspace{5pt} At this point, let us make some general comments about the concept of central charge. We are interested in phase transitions which are described by conformal field theories\cite{CFT} --a situation not uncommon in $1+1$ dimensions. These encode the long-distance physics of the model, such as the asymptotics of correlation functions, and they are characterized by certain universal numbers. One of the most important numbers, relevant to all CFTs, is the central charge $c>0$. It is sometimes said to be a measure of the gapless degrees of freedom. There are at least three reasons for that. Firstly, for small but finite temperatures $T$, the specific heat $C$ is linearly proportional to the central charge $c$, more precisely $C \propto c T$. Secondly, if one stacks two decoupled CFTs, the central charge is additive. Thirdly, there is Zamolodchikov's $c$-theorem\cite{Zamolodchikov1986}, which says that under renormalization group flows, the central charge can only decrease. The latter is consistent with the idea that renormalization removes (high-energy) degrees of freedom. The central charge of a CFT is one of its most crucial pieces of information. In fact under certain conditions of unitarity and minimality, all CFTs with $0 < c < 1$ have been classified, and for any such $c$ there are only a finite number of CFTs possible\footnote{More precisely, $0<c<1$ fixes a finite list of \emph{possible} scaling dimensions for so-called primary operators. Which subset of these dimensions are realized in any particular CFT can vary.}. On the other hand, while a lot is known about CFTs with $c\geq 1$, it is not known how many exist and which numbers characterize them.

\textbf{Transition between SPT and trivial phase} \hspace{5pt} For clarity, let us state the conjecture now and give some conceptual motivation. In section \ref{sec:alpha_transition} we then illustrate how would one naturally arrive at this conjecture by investigating transitions in the $\alpha$-chain model, both with and without interactions.

\hspace{10pt}

\emph{\textbf{Conjecture:} Consider the transition between a trivial phase and an SPT phase with a $d$-dimensional protected edge mode (on each edge). If the transition is described by a CFT, then its central charge $c$ is lower bounded by $\log_2 d$.}

\hspace{10pt}

The intuitive picture is the following. An SPT phase has well-defined edge modes which are localized up to the correlation length $\xi$. As long as the edges cannot communicate with one another and the relevant symmetries are preserved, then the modes cannot be gapped out. Hence there are only two ways of trivializing the system: either there is a discontinuous change (signaling a first order transition) or the edge modes become delocalized such that those of the left and right edge can overlap and hybridize. The latter requires $\xi \to \infty$, corresponding to a continuous transition. Hence at the transition we expect the delocalized edge modes and their long-wavelength fluctuations to become the bulk gapless fields. Since $d$ is a measure of the former whereas the central charge $c$ is a measure of the latter, we arrive at a relationship. Note that the logarithm ensures this bound is additive when considering two decoupled chains, similar to the central charge. Moreover, if one introduces a coupling between two such chains, $d$ can only decrease, again similar to $c$. One can only give a lower bound since there might always be extra gapless fields present.

\textbf{Transition between different SPT phases} \hspace{5pt} On first sight, allowing transitions between two non-trivial SPT phases seems a more complicated problem. We now argue how this is not the case, by using the group structure\cite{Fidkowski-2010,Turner-2010,Chen-2010} of SPT phases. Suppose one has a path in parameter space between SPT A and SPT B, possibly with critical points along the way. Everywhere along this line, one can stack with the inverse of SPT B. Note that due to SPT B being gapped, this cannot affect a CFT describing a critical point, although now the gapped phases have been reduced to the previous case. Hence:

\hspace{10pt}

\emph{\textbf{Corollary:} Consider the transition between two SPT phase characterized by symmetry fractionalizations $\rho_A$ and $\rho_B$. If it is described by a CFT, then its central charge $c$ is lower bounded by $\log_2 \dim \rho_A\rho_B^{-1}$ (where $\dim$ represents the quantum dimension for a single edge).}

\hspace{10pt}

In section \ref{sec:alpha_transition} we show how to arrive at the above conjecture by investigating (free and interacting) transitions between the SPT phases we have discussed earlier in this paper. Moreover this leads to certain predictions for interacting phase diagrams, which we confirm by DMRG. In section \ref{sec:golden} we test our conjecture for the critical points between so-called golden chain SPT phases --generalizations of the Kitaev chain-- and the trivial phase, which in fact realize \emph{all} unitary minimal CFTs with $0 < c < 1$. Some examples of known topological transitions for bosonic systems, including Wess-Zumino-Witten models, are discussed in section \ref{sec:other_transitions}.

\subsection{Transitions between the $\alpha$-chains: free and interacting} \label{sec:alpha_transition}

\textbf{The $\boldsymbol{c=\frac{1}{2}}$ CFT} \hspace{5pt} The critical point between the Kitaev chain and the trivial chain is well-known to be described by a non-chiral massless Majorana field in the continuum limit\cite{CFT,Cardy86c,Boyanovsky89}. This defines the unique\footnote{It is unique if one counts local \emph{and} non-local primary fields together, e.g. see (7.20) in Ginsparg's notes\cite{Ginsparg}. (For some boundary conditions, non-local fields contribute to the finite-size spectrum.) If one chooses the labeling `local' versus `non-local' to be part of the CFT data, there are multiple $c=\frac{1}{2}$ CFTs (see main text).} unitary CFT with central charge $c=\frac{1}{2}$. Aside from $c$, other important information characterizing a CFT is the list of scaling dimensions (specifying the power-law decay of correlation functions) of so-called primary fields (generating all other fields of the theory). The $c=\frac{1}{2}$ CFT has five such non-trivial primaries ($\sigma,\mu,\psi,\bar \psi,\varepsilon$ with respective scaling dimensions $\frac{1}{8},\frac{1}{8},\frac{1}{2},\frac{1}{2},1$)\cite{Ginsparg}. In this (fermionic) realization of the CFT, the \emph{local} operators are the Majorana fields $\psi,\bar{\psi}$ and the mass term $\varepsilon$, whereas the \emph{non-local} $\sigma$ and $\mu$ are string order parameters for, respectively, the nearby topological and trivial phases.

Under Jordan-Wigner, we map to the critical Ising chain. This is described by the same CFT, but --similar to the gapped case-- the non-local transformation has changed the physics\cite{Cardy86c,Boyanovsky89}. In particular, the Ising order parameter field $\sigma$ is now local. Nevertheless, the central charge is unchanged. In fact, one can argue in elementary terms that the Jordan-Wigner transformation \emph{always} preserves the central charge. Indeed, it does not change the entanglement structure in the computational basis, and for a CFT this fixes the central charge\cite{Calabrese04}. More exactly, if $S$ is the entanglement between a region of size $L$ and the rest of the system, then $c = \lim_{L\to \infty} \frac{3 \; S}{\log L}$.

\textbf{Transitions between two different $\bm \alpha$-chains} \hspace{5pt} If one interpolates between the $\alpha_0$-chain and the $\alpha_1$-chain, $H = (1-\lambda) H_{\alpha_0} + \lambda H_{\alpha_1}$, then it is straightforward to show that the single-particle spectrum is gapped everywhere, except at $\lambda = \frac{1}{2}$, where the single-particle spectrum has $|\alpha_0-\alpha_1|$ linear crossings through the Fermi surface. Each crossing can be linearized to give an effective Majorana field. In other words, the CFT is a direct sum of $|\alpha_0-\alpha_1|$ copies of a single Majorana CFT. In particular $c = \frac{|\alpha_0-\alpha_1|}{2}$. The fact that this only depends on the difference is consistent with the argument we gave for the transition between any two SPT phases being reducible to the transition between an SPT phase and the trivial phase. The fact that the transition is described by a stack of Majorana CFTs is also intuitive from Fig.~\ref{fig:alpha}: for example, $H = H_0 + H_2$ can be pulled apart into two decoupled critical Kitaev chains. Note that in case $\alpha_1 = 0$, the above can be rewritten as $c = \log_2 d$, since the non-interacting $\alpha_0$-chain has $d$-dimensional protected edge modes with $d = 2^{|\alpha_0|/2}$.

\textbf{Transitions between generalized cluster models} \hspace{5pt} Unlike the fermionic case, the bosonic critical theories are not simply stacks of a single critical chain. To be more precise, the \emph{local} primaries are not just obtained from the \emph{local} primaries of a single chain. This makes physical sense: the reason for the critical theories on the fermionic side being stacks, is due to the (additive) group structure of SPT phases. In section \ref{scn:sSPT} we already saw how this group structure is not preserved under the Jordan-Wigner transformation. More concretely, one should not expect the topological transition between the cluster model and the trivial phase to be described by a stack of symmetry breaking Ising transitions. Instead, the transition between two generalized cluster model with respective $\alpha = \alpha_0$ and $\alpha = \alpha_1$ is naturally described by the \emph{bosonized} description of $|\alpha_0-\alpha_1|$ massless Majorana fields. This is referred to as the Wess-Zumino-Witten (WZW) $SO(|\alpha_0-\alpha_1|)_1$ field theory with central charge $c = \frac{|\alpha_0-\alpha_1|}{2}$. Lahtinen et al.\cite{SON} performed finite-size scaling on these spin models and found perfect consistency with the field theory predictions. From now on we will only focus on the central charge of the transitions, hence we can go back and forth between the bosonic and fermionic language without any further disclaimers.

\begin{figure}[h]
	\includegraphics[scale=.75]{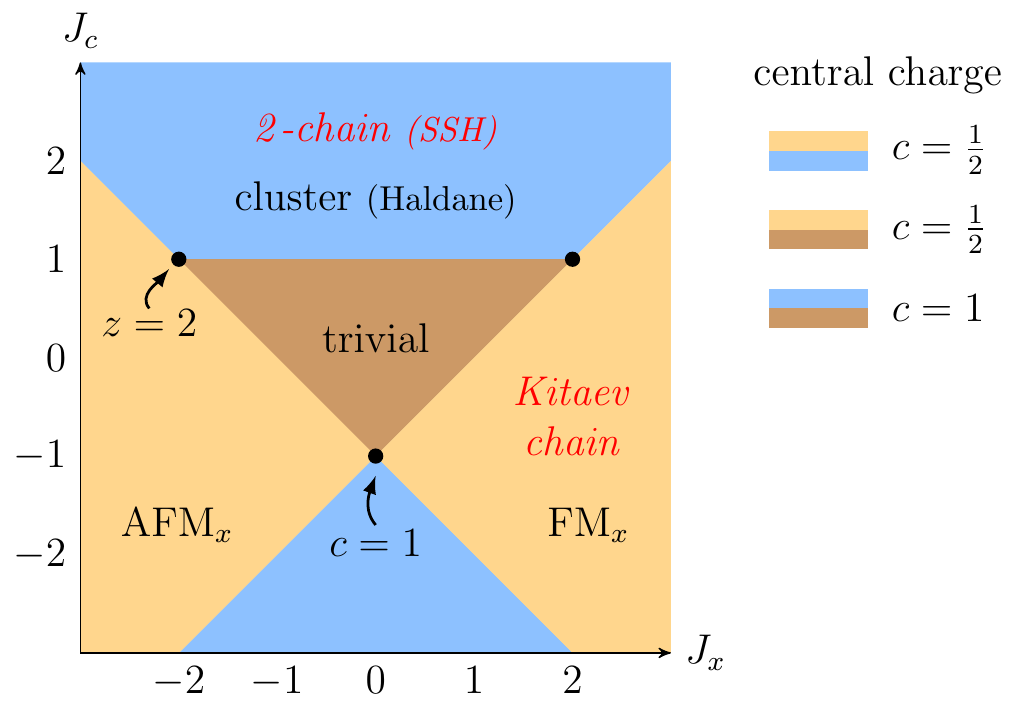}
	\caption{Phase diagram for the non-interacting Hamiltonian \eqref{ham:phase_diagram}\label{fig:phase_diagram_free}}
\end{figure}

\textbf{Phase diagrams in the absence of interactions} \hspace{5pt} So far we have only discussed the transitions that arise due to the linear interpolation of any two $\alpha$-chains. Let us briefly discuss the case of multi-parameter phase diagrams. The key point is that the intuition from the one-dimensional phase diagrams discussed before, extends to the more general case. Let us for example look at the phase diagram for
\begin{align} \label{ham:phase_diagram}
H_\textrm{ferm} &= H_0 + J_1 H_1 + J_2 H_2 \; \textrm{, or} \\
H_\textrm{spin} &= \sum_n \left( Z_n - J_1 X_n X_{n+1} - J_2 X_{n-1} Z_n X_{n+1} \right) \notag
\end{align}
Note that these two Hamiltonians map to each other under the Jordan-Wigner transformation, hence they have the same phase diagram and the same central charges at the transitions. The analytical result is shown in Fig.~\ref{fig:phase_diagram_free}, where the labels are in black for the spin variables, and in red for the fermionic ones (without repeating `trivial', which is the same in both variables). We recognize the three phases corresponding to the three Hamiltonians that appear in Eq.~\ref{ham:phase_diagram}, i.e. the trivial phase, the Kitaev chain and the $2$-chain. More importantly, we see the central charges exactly correspond to the $\log_2 d$ formula. For example, by the group structure of SPT phases the transition between the $2$-chain phase and the Kitaev phase should be the same as the Kitaev phase to the trivial phase, for which we predict $c = \log_2 d = \log_2 \sqrt{2} = \frac{1}{2}$.

The more general insight is that if one starts from a gapped phase which is adiabatically connected to the $\alpha_0$-chain, then the phase transition to a gapped phase connected to the $\alpha_1$-chain will generically have a central charge $c = \frac{|\alpha_0-\alpha_1|}{2}$. Similar phase diagrams have been obtained before\cite{Montes12,Niu12,SON,Ohta16}. We will be interested in what happens to such phase diagrams in the presence of interactions.

\textbf{The effect of interactions on the transitions} \hspace{5pt}
Let us now consider the effect of interactions when going from one $\alpha$-chain to the other. In fact, some of the interacting transitions between different stacks of Kitaev chains were already discussed in the seminal work of Fidkowski and Kitaev\cite{Fidkowski-2010}. In particular they discussed how in the non-interacting case, the transition between the $4$-chain and the trivial phase has $c = 2$ --as we arrived at above-- which in the presence of interactions reduces to $c=1$. This is natural from section \ref{sec:Hubbard} where we identified the interacting $4$-chain with the alternating spin-$\frac{1}{2}$ Heisenberg chain. More precisely, it shows that for strong interactions, the phase transition between the $4$- and $0$-chain is exactly given by the spin-$\frac{1}{2}$ Heisenberg chain. This well-studied model is known to have $c=1$ (more completely it is described by WZW $SU(2)_1$, or equivalently as a Luttinger liquid with $K = \frac{1}{2}$). At the same time, we know that for the gapped $4$-chain itself, interactions reduce the edge degeneracy from $d=4$ to $d=2$. This suggests that the degrees of freedom at the transition are linked to the degrees of freedom on the edge in the gapped SPT phase.

Having looked at $\alpha = 4$, we now consider the cases of $\alpha=1,2,3$ (which by the symmetries discussed in section \ref{scn:fSPT} cover all the cases of the $\mathbb Z_8$ inequivalent SPT phases). One does not expect the central charges of the transitions from the Kitaev chain or the $2$-chain to the trivial phase to change, as these CFTs are well-known to be stable against interactions. Less is known about the CFT describing the transition between the $3$-chain and the trivial chain, which in the free case has $c = \frac{3}{2}$. However, since interactions do \emph{not} affect the three Majorana modes on its left edge, i.e.\footnote{Note that a non-integer $d$ is to be interpreted as ``if one has $n$ such edges, it asymptotically has the Hilbert space dimension $d^n$''.} $d = 2\sqrt{2}$ (such that the total degeneracy is $d^2 = 8$), one might expect its transition also to remain unchanged. To test this hypothesis, let us consider the transition between the Kitaev chain and the interacting $4$-chain (which by the group structure of SPTs also describes the transition between the $3$-chain to the trivial phase):
\begin{equation} \label{ham:HaldaneKitaev}
H_\textrm{ferm} = H_4 + J_1 H_1 + V
\end{equation}

\begin{figure}[h]
	\includegraphics[scale=.8]{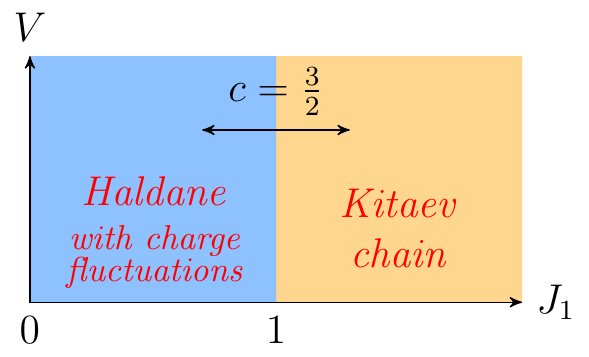}
	\caption{Phase diagram for the interacting Hamiltonian \eqref{ham:HaldaneKitaev}\label{fig:KitaevHaldane}}
\end{figure}

Here $V$ is the interaction term introduced in Eq.~\eqref{4chain}. As discussed in section \ref{sec:Hubbard}, $H_4+V$ is equivalent to the alternating spin-$\frac{1}{2}$ Heisenberg chain with charge fluctuations. Fig.~\ref{fig:KitaevHaldane} shows the resulting phase diagram we obtain using infinite DMRG (iDMRG)\cite{Kjall13}, where the central charge is extracted from entanglement scaling\cite{Calabrese04,Tagliacozzo,Pollmann-2009}: the system is tuned to criticality, where for each bond dimension $\chi$ there is an optimal infinite matrix product state approximating the ground state with an effective correlation length $\xi$ and bipartition entanglement $S$, obeying the scaling relationship $S=\frac{c}{6}\log \xi$. Note that one can define the duality transform $\gamma_n \to \gamma_{5-n}$ and $\tilde \gamma_n \to \tilde \gamma_{-n}$, which switches $H_1 \leftrightarrow H_4$ and leaves $V$ invariant. Hence the critical line shown in Fig.~\ref{fig:KitaevHaldane} exactly corresponds to the self-dual coupling $J_1 = 1$, which is useful for entanglement scaling since one can exactly tune to the transition.

We summarize our findings in Table~\ref{table:interacting_transitions}. We see that in all these cases the central charge at the transition is given by the number of Majorana modes in the SPT phase, weighted by a factor of $\frac{1}{2}$. More concisely, this is $\log_2 d$, where $d$ is the degeneracy of a single edge. We expect this relationship to hold for the transitions between these phases, even with more complicated Hamiltonians. However, more generally, we can have models where $\log_2 d$ is not even rational, whereas the central charge always is (at least for the CFTs so far encountered in these contexts). Nevertheless we will see that even in those cases, $\log_2 d$ provides a lower bound, which in many cases is in fact very close to the true value of $c$.

\begin{table}[h]
	\begin{tabular}{c|c|c||c|c}
		\multicolumn{1}{c}{} & \multicolumn{2}{c||}{\textbf{free}} & \multicolumn{2}{c}{\textbf{interacting}} \\
		\cline{2-5} $\alpha$ & central charge $c$ & $\log_2 d$ & central charge $c$ & $\log_2 d$ \\ 
		\hline $-3 \leq \cdots \leq 3$ & $\frac{|\alpha|}{2}$ & $\frac{|\alpha|}{2}$ & $\frac{|\alpha|}{2}$ & $\frac{|\alpha|}{2}$ \\
		$4$ & $2$ & $2$ & $1$ & $1$
		\end{tabular}
	\caption{Phase transitions from the $\alpha$-chain to the trivial chain. The first set of columns give the central charge and the degeneracy $d$ of a single edge in the absence of interactions. The second set of columns is in the presence of $T$-preserving interactions.}
	\label{table:interacting_transitions}
\end{table}

\begin{figure}[h]
	\includegraphics[scale=.75]{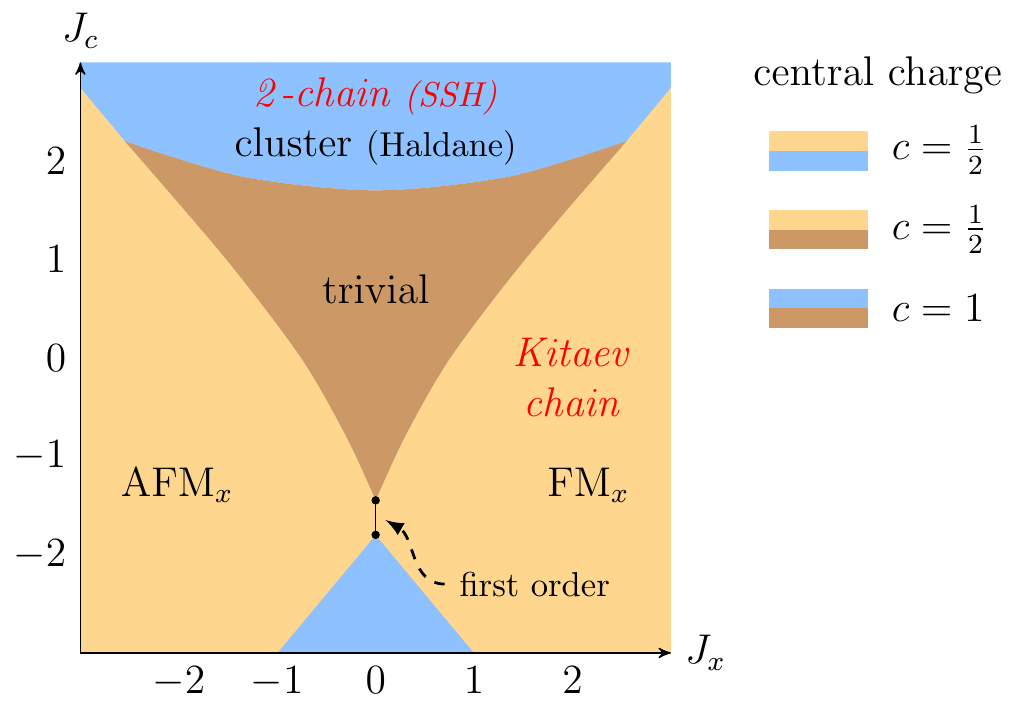}
	\caption{Phase diagram for the Hamiltonian \eqref{ham:phase_diagram} in the presence of the interactions described in the main text \label{fig:phase_diagram_int}}
\end{figure}

\textbf{Implications for interacting phase diagrams} \hspace{5pt} Before testing the conjecture that $c \geq \log_2 d$ for other types of SPT phases, we first check its validity in more complicated phase diagrams than those involving direct interpolations between two (possibly interacting) $\alpha$-chains. Let us first consider an interacting version of Hamiltonian \eqref{ham:phase_diagram}, adding an interaction which in spin language corresponds to $\frac{1}{2}\sum_n  Z_nZ_{n+1}$ or in fermionic variables $\frac{1}{2}\sum_n \tilde \gamma_n \tilde \gamma_{n+1} \gamma_n \gamma_{n+1}$. The resulting phase diagram in Fig.~\ref{fig:phase_diagram_int} was mapped out using iDMRG, identifying each phase in terms of its entanglement properties\cite{Pollmann12b}. The central charge at the critical points was extracted using entanglement scaling\cite{Calabrese04,Tagliacozzo,Pollmann-2009} as explained above. We see that we obtain the central charges we expect based on the protected edge degeneracies. In addition a first order transition appears between the two Ising phases which are only distinguished by (un)broken translation symmetry.

\begin{figure}[h]
	\includegraphics[scale=.8]{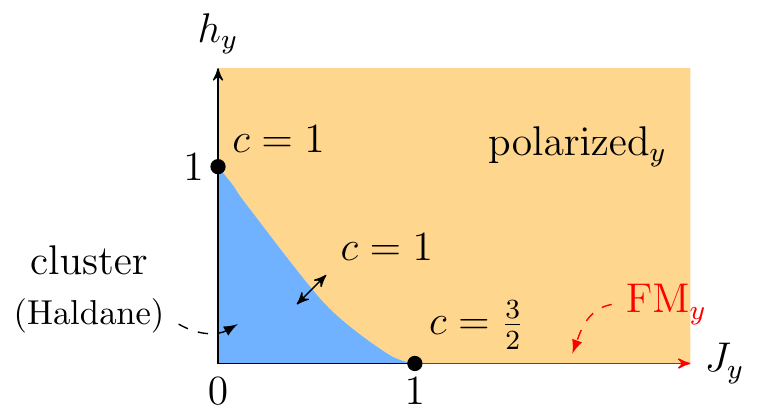}
	\caption{Phase diagram for the spin model \eqref{ham:spin_model} which does not correspond to any fermionic model if $h_y \neq 0$. The central charges can be predicted from our conjecture and are confirmed by iDMRG.\label{fig:phase_diagram_spin}}
\end{figure}

Secondly, let us consider the spin Hamiltonian
\begin{equation}
H = \sum_n \left( -X_{n-1}Z_n X_{n+1} - J_y Y_n Y_{n+1} + h_y Y_n \right)
\label{ham:spin_model}
\end{equation}
For $h_y = 0$ this is equivalent to the free-fermion model which interpolates between the $2$-chain and the $(-1)$-chain, hence $c=\frac{3}{2}$ at the transition. However for $h_y \neq 0$ the model has no $\mathbb Z_2$ symmetry and is hence not dual to \emph{any} fermionic model. In this case we only have our conjecture to fall back on. Since $h_y \neq 0$ explicitly breaks the symmetry which was originally spontaneously broken for large $J_y$, we can conclude that the large $J_y$ phase is now trivial. On the other hand for small $J_y$ (and $h_y$) we are in the cluster phase, which is still a well-defined SPT phase protected by $PT = \left(\prod_n Z_n\right) K$. In particular each edge mode has a degeneracy $d=2$, such that for $h_y \neq 0$ we expect that at the critical point $c=\frac{3}{2}$ reduces to $c=\log_2 2 = 1$. This is confirmed in Fig.~\ref{fig:phase_diagram_spin} which was obtained using the methods described in the previous paragraph.

\subsection{All minimal CFTs as transitions between SPT phases} \label{sec:golden}

Here we test our conjecture for two \emph{known} types of generalizations of the Kitaev chain: firstly for so-called parafermionic chains, and secondly for anyonic chains with $SU(2)_k$ statistics (also called golden chains). In the latter case, the critical theories describing the phase transition to the trivial phase in fact capture all unitary minimal CFTs with $0<c<1$. Our conjectured bound is confirmed in each case, which moreover only underestimates $c$ by less than one percent.

\textbf{Parafermions} \hspace{5pt} Instead of Majorana modes, one can consider $(N\geq 2)$-parafermionic operators $\gamma_a$ which satisfy
\begin{equation}
\gamma_a^\dagger = \gamma_a^{N-1} \; , \quad \gamma_a^N = 1 \quad \textrm{and} \quad \gamma_a \gamma_b = e^{i2\pi/N} \gamma_b \gamma_a
\end{equation}
For $N=2$ we recover the Majorana algebra. Analogously to the Kitaev chain, these can form an SPT phase\cite{Fendley12} with an edge degeneracy $d = \sqrt{N}$. There is only a direct, second order transition to a trivial phase when $N = 2,3,4$, described by so-called parafermion CFTs\cite{ParafermionCFT}. Their central charges are summarized in Table~\ref{table:parafermion}, confirming our conjecture. The last column shows the difference between the central charge and our lower bound. Curiously, for $N=3$ we have $c = \frac{4}{5} = 0.8$ and $\log_2 d = \log_2 \sqrt{3} \approx 0.7925$, such that our lower bound is saturated within one percent.

\begin{table}[h]
	\begin{tabular}{c|c|c|c}
		$N$ & central charge $c$ & $\log_2 d$ & $\frac{c-\log_2 d}{c}$\\ 
		\hline $2$ & $\frac{1}{2}$ & $\frac{1}{2}$ & $0$ \\
		$3$ & $\frac{4}{5}$ & $\log_2 \sqrt{3}\approx 0.7925$ & $\approx 0.0094$ \\
		$4$ & $1$ & $1$ & $0$
	\end{tabular}
	\caption{The central charges for the $N$-parafermion CFTs at the critical point between trivial phase and SPT phase with edge mode $d=\sqrt{N}$. Comparison to the conjectured lower bound $\log_2 d$. Expressions for $c$ obtained from Ref.~\onlinecite{ParafermionCFT} and for $d$ from Ref.~\onlinecite{Fendley12}.}
	\label{table:parafermion}
\end{table}

\textbf{Golden chains: Fibonacci and $\bm{SU(2)_k}$ anyons} \hspace{5pt} A different generalization of Majorana modes is obtained by interpreting them as non-abelian anyons obeying the $SU(2)_2$ fusion rule\footnote{The $SU(2)_k$ fusion rules on $0,\frac{1}{2},\cdots,\frac{k}{2}$ are given by $j_1 \times j_2 = |j_1-j_2| + \cdots + \min{}(j_1+j_2, k-j_1-j_2)$, so for $k=2$ we have $\frac{1}{2}\times \frac{1}{2} = 0 + 1$ and $1 \times 1 = 0$. Here $\frac{1}{2}$ is identified with the Majorana anyon and $0$ ($1$) with an empty (filled) fermionic mode.} $\gamma \times \gamma = 1 + \psi$, i.e. two Majorana modes define a fermionic mode which can be empty ($1$) or filled ($\psi$). For any $k \geq 2$, one can consider non-abelian anyons obeying a so-called $SU(2)_k$ fusion rule, and analogously to the Kitaev chain they can form an SPT phase\cite{DeGottardi14} where the edge mode has the quantum dimension\footnote{E.g.: for the topological phase of Fibonacci anyons, stacking $N$ open chains gives rise to a $F_{2N+1}$-fold degeneracy (where $F_n$ denotes the $n$-th Fibonacci number).} of the underlying anyons, given by the Beraha numbers $d=\sqrt{B_{k+2}}=2 \cos \frac{\pi}{k+2}$. (These models are referred to as golden chains, since for $k=3$ we obtain Fibonacci anyons where the quantum dimension equals the golden ratio $\varphi = 2 \cos \frac{\pi}{5} = \frac{1+\sqrt{5}}{2}$.) For each $k$ there is a direct continuous transition to the trivial phase\cite{fibonacci}. Interestingly these transitions give rise to all central charges $0<c<1$ possible for unitary minimal CFTs. Our lower bound is confirmed in each case, and we again find that it captures the true value of $c$ within one percent. The situation is summarized in Table.~\ref{table:golden}.

\begin{table}[h]
	\begin{tabular}{c|c|c|c}
		$k$ & central charge $c$ & $\log_2 d$ & $\frac{c-\log_2 d}{c}$\\ 
		\hline $2$ & $\frac{1}{2}$ & $\frac{1}{2}$ & $0$ \\
		$3$ & $\frac{7}{10}$ & $\approx 0.6942$ & $\approx 0.0082$ \\
		$4$ & $\frac{4}{5}$ & $\approx 0.7925$ & $\approx 0.0094$ \\
		$5$ & $\frac{6}{7}$ & $\approx 0.8495$ & $\approx 0.0089$ \\
		$\vdots$ & $\vdots$ & $\vdots$ & $\vdots$ \\
		$k$ & $1-\frac{6}{(k+1)(k+2)}$ & $\log_2 \left( 2 \cos \frac{\pi}{k+2}\right)$ & $0\leq \frac{c-\log_2 d}{c} <\frac{1}{100}$ \\
		$\vdots$ & $\vdots$ & $\vdots$ & $\vdots$ \\
		$\infty$ & $1$ & $1$ & $0$
	\end{tabular}
	\caption{The central charges for the unitary minimal CFTs at the critical point between trivial phase and SPT phase of $SU(2)_k$ anyons with edge mode $d=2 \cos \frac{\pi}{k+2}$. Comparison to the conjectured lower bound $\log_2 d$. Expressions for $c$ and $d$ obtained from Ref.~\onlinecite{fibonacci}.}
	\label{table:golden}
\end{table}

\subsection{Testing the conjecture at known bosonic SPT transitions} \label{sec:other_transitions}

In this section we review some known phase transitions between bosonic SPT phases and the trivial phase, and compare their central charges to our conjectured lower bound. Firstly we focus on the case with discrete symmetries, and afterwards on $SU(N)$ spin chains. It is worth noting that work on the former has led to a different constraint on the central charge of a phase transition between SPT phases, which is also a corollary of our conjecture, as we will discuss.

\textbf{SPT phases protected by $\bm{\mathbb Z_n \times \mathbb Z_n}$} \hspace{5pt} In recent work by Tsiu et al.\cite{Tsui17}, bosonic SPT phases protected by a $\mathbb Z_n \times \mathbb Z_n$ symmetry were constructed. These have degenerate edge modes with $d = n$, and there is a direct continuous transition to the trivial phase if $n=2,3,4$. These transitions obey our conjecture, as shown in Table~\ref{table:ZnxZn}.

\begin{table}[h]
	\begin{tabular}{c|c|c|c}
		$n$ & central charge $c$ & $\log_2 d$ & $\frac{c-\log_2 d}{c}$\\ 
		\hline $2$ & $1$ & $1$ & $0$ \\
		$3$ & $\frac{8}{5}$ & $\log_2 3 \approx 1.585$ & $\approx 0.0094$ \\
		$4$ & $2$ & $2$ & $0$
	\end{tabular}
	\caption{The central charges for the CFTs at the critical point between trivial phase and SPT phase protected by $\mathbb Z_n \times \mathbb Z_n$ with edge mode dimension $d=n$. Comparison to the conjectured lower bound $\log_2 d$. Expressions for $c$ and $d$ obtained from Ref.~\onlinecite{Tsui17}.}
	\label{table:ZnxZn}
\end{table}

Moreover, in the same article it is proven that the critical point between \emph{any} bosonic SPT phase and the trivial phase must always have a central charge $c \geq 1$. This also follows from our conjecture, since a bosonic SPT phase can only have an integer dimension for its edge mode (to see this, note that the edge mode of a bosonic SPT phase transforms under a projective representation of a symmetry group). Hence $d \geq 2$ such that $c \geq \log_2 d \geq 1$.

\textbf{$\bm{SU(N)}$ spin chains} \hspace{5pt} In the work of Nonne et al.\cite{Nonne13}, SPT phases protected by $SU(2M)$ --or more correctly\cite{Duivenvoorden13}, $PSU(2M)$-- were constructed. These are natural generalizations of the AKLT model (which corresponds to $M =1$), where each edge has a degeneracy $d =\frac{(2M)!}{M!M!}$. The natural expectation for the critical theory describing the phase transition to the trivial phase, is the so-called Wess-Zumino-Witten (WZW) field theory for the group $SU(2M)$ with respect to some level $k =1,2,\dots$, referred to as WZW $SU(2M)_k$. It is sufficient to verify our lower bound for the case $k=1$, since this has the smallest central charge, with $c = 2M-1$. This is a non-trivial check of our conjecture, since $d$ blows up exponentially with $M$. One can use the Stirling approximation to show that $\log_2 d \leq 2M - \frac{1}{2} \log_2 M$ for \emph{all} $M > 0$ (which also gives the asymptotic expression for large $M$). This indeed lower bounds the central charge if $M \geq 4$. The remaining cases $M=1,2,3$ can be checked by hand, as shown in Table~\ref{table:SU(N)}. Note that the above Stirling approximation also shows the relative difference between $c$ and $\log_2 d$ goes to zero as $M \to \infty$.

Ref.~\onlinecite{Nonne13} did not discuss the transition from the above SPT phase to the trivial phase. Instead it considered the transition to a spontaneously dimerized phase, which was suspected to be described by WZW $SU(2M)_2$ --in direct generalization of the case of the AKLT model. If one would explicitly break translation symmetry, then there is either a direct transition to the trivial phase, or a new intermediate phase. In case of the former, one would expect on general grounds\cite{Furuya17,Lecheminant15} that the critical point would flow to WZW $SU(2M)_1$ --extending the case of the AKLT model\cite{Kitazawa99}. Since it is known\cite{Nataf16} that obtaining reliable entanglement scaling for models with these large symmetry groups requires large-scale numerics explicitly incorporating the non-abelian symmetries, we do not attempt a numerical verification of this here.

\begin{table}[h]
	\begin{tabular}{c|c|c|c}
		$M$ & central charge $c$ & $\log_2 d$ & $\frac{c-\log_2 d}{c}$\\ 
		\hline $1$ & $1$ & $1$ & $0$ \\
		$2$ & $3$ & $\log_2 6 \approx 2.59$ & $\approx 0.14$ \\
		$3$ & $5$ & $\log_2 20 \approx 4.32$ & $\approx 0.14$ \\
		$4$ & $7$ & $\log_2 70 \approx 6.13$ & $\approx 0.12$ \\
		$\vdots$ & $\vdots$ & $\vdots$ & $\vdots$ \\
		$M$ & $c = 2M -1$ & $\log_2 \frac{(2M)!}{(M!)^2}$ & $\sim \frac{\log_2 M}{4M}$ ($M$ large) \\
		$\vdots$ & $\vdots$ & $\vdots$ & $\vdots$ \\
		$\infty$ & $\infty$ & $\infty$ & $0$
	\end{tabular}
	\caption{The central charges for WZW $SU(2M)_1$ which likely describe the critical point between trivial phase and the SPT phase protected by $SU(2M)$ with edge mode $d=\frac{(2M)!}{(M!)^2}$ (see main text). Comparison to the conjectured lower bound $\log_2 d$. Expression for $d$ obtained from Ref.~\onlinecite{Nonne13}.\label{table:SU(N)}}
\end{table}

More generally, it is known that there are $N-1$ distinct topological phases protected by $PSU(N)$ symmetry \cite{Duivenvoorden13}. Except for the aforementioned case, these are all \emph{chiral} if $N>2$ --i.e. they are not left-right symmetric. For a given $N$, all these phases can be generated by stacking copies of a chiral chain with an edge mode on the left (right) transforming under the fundamental (conjugate) representation of $SU(N)$ --although inter-chain couplings are needed to remove accidental degeneracies. A Hamiltonian for such a single generating chain (with $d=N$) was constructed by Roy and Quella\cite{Roy2015-preprint} and the transition to the trivial phase (in the form of a dimerized phase with \emph{explicit} translation symmetry breaking) was argued to be described by WZW $SU(N)_1$ (with $c=N-1$)\cite{Roy2015-preprint,Lecheminant15}. Again our conjecture is confirmed, although now our lower bound $\log_2 d$ is much less tight --it only grows logarithmically with $N$, whereas $c$ grows linearly. We note that the analysis of Roy and Quella agrees with complementary approaches, such as the work by Rao et al.\cite{Rao16} reporting an SPT phase with $SU(3)$ symmetry and a phase transition to the trivial phase described by WZW $SU(3)_1$.

\sectionheading{Conclusion}\label{scn:conclusion}

In the first part of this work, we showed how various SPT models can be related to the $\alpha$-chain by using symmetries as a guide. This gives a unifying picture of known models, identifying the SSH model with a stack of two Kitaev chains, and the cluster model with a close cousin of the AKLT chain. These two set of models moreover map to each other by the non-local Jordan-Wigner transformation, which more generally relates the $\alpha$-chain to the generalized cluster models. This offers several open questions, such as whether the emergent $SO(3)$ symmetry we saw for the cluster model generalizes to other values of $\alpha$, and whether the generalized cluster models are also connected to shorter-range higher-spin models (like the cluster model being adiabatically connected to the alternating spin-$\frac{1}{2}$ Heisenberg chain, which on its turn connects\cite{Hida92} to the spin-$1$ Heisenberg chain).

Our approach shed light on the topological Hubbard chain which connects the stack of four Kitaev chains to a spin chain in the Haldane phase. This model illustrates that if we reinterpret the Haldane phase to be protected by for example fermionic sublattice symmetry, rather than time-reversal symmetry, it is stable against charge fluctuations. This constructions also leads to a simpler path from the $8$-chain to the trivial phase. It is an interesting issue whether this symmetry-guided approach can be applied more generally to the program laid out in Ref.~\onlinecite{You2015b}, where fermionic SPT phases are understood in terms of bosonic ones.

In the second part, we studied the phase transitions between SPT phases, in particular leading to the conjecture that the central charge $c$ at the transition between the trivial phase and an SPT phase with edge modes of dimension $d$ is lower bounded by $\log_2 d$. This opens up a number of important questions. Firstly, it is desirable to better understand the curious relationship we found between the central charge $c_m$ of the minimal model $\mathcal M(m+1,m)$ (for any $m \geq 2$) and the Beraha\cite{Saleur90} numbers, namely $4^{c_m} \approx B_{m+1}$. Secondly, aside from numerically studying the phase transitions which we discussed in the context of the PSU(N) spin chains in section \ref{sec:other_transitions}, it would be interesting to check our conjecture for the $SO(2M+1)$ SPT phases defined in Ref.~\onlinecite{Tu08}. Since these have edge modes with dimension $d=2^M$, we obtain the non-trivial conjecture that $c \geq M$ when transitioning to the trivial phase. In fact, the transition to a dimerized phase is known\cite{Alet11,Tu11} to have $c = M + \frac{1}{2}$, such that the $c$-theorem\cite{Zamolodchikov1986} suggests an upper bound for the transition to the trivial phase, obtaining the tight condition $M \leq c \leq M +\frac{1}{2}$. More generally, if no counter-example to our conjecture is to be found, it would be very valuable to find a proof --likely offering insights into the structure of CFTs. In particular, it would offer a formalization of the intuitive idea that the central charge measures the relevant degrees of freedom. Moreover, it would constitute the first step towards a general understanding of topological phase transitions in one dimension, whose concepts might aid the same task in higher dimensions.

\sectionheading{Acknowledgements}

The authors would like to thank Nick Jones, Siddhard Morampudi, Ari Turner, Ville Lahtinen, Curt von Keyserlingk, Paul Fendley, Hong-Hao Tu, Max Metlitski, Zheng-Cheng Gu, Tibor Rakovszky and Henrik Dreyer for helpful discussions. RV was supported by the German Research Foundation (DFG) through the Collaborative Research Centre SFB 1143 and FP acknowledges support from DFG through Research Unit FOR 1807 with grant no. PO 1370/2-1 and from the Nanosystems Initiative Munich (NIM) by the German Excellence Inititiative. This research was partly performed while RV was visiting the Perimeter Institute for Theoretical Physics.

\bibliography{1D_SPT.bbl}
%\bibliography{bibo}

\appendix
\section{The principles of symmetry fractionalization in 1D} \label{app:sym}
Consider a gapped one-dimensional system of length $N$ invariant under a global symmetry group $G$. The total Hilbert space has a tensor product structure $\mathcal H = \otimes_n \mathcal H_n$ with an on-site Hilbert space dimension dim$(H_n) =d$ (possibly after blocking). The abstract symmetry group $G$ acts via an (anti)linear representation $\rho: G \to U(d^N)$ on the Hilbert space, where $U\left(d^N\right)$ are the $d^N\times d^N$ unitary matrices. We work in the setting where the symmetry is on-site, which means that there exists an (anti-)linear representation $\rho_n: G \to U(d)$ such that for all $g \in G$, $\rho(g) = \otimes_n \rho_n(g)$. In the case of an anti-unitary symmetry, this means that the basis in which we define complex conjugation has to be compatible with the tensor product structure. Note that such on-site symmetries are automatically well-defined if we have open boundary conditions, which is essential for our approach. Since we will be interested in faithful representations (which means $G \cong \rho(G)$), we will in fact identify $G$ with its representation. In other words we can say `take $U \in G$' where $U$ is some unitary operator.

\textbf{Each symmetry fractionalizes} \hspace{5pt} Consider a bosonic system with open boundaries. In section \ref{scn:SPT} we have argued that any unitary symmetry $U\in G$ can effectively be written as $U = U_L U_R$ where $U_{L,R}$ are exponentially localized near the boundary. This means that in the thermodynamic limit, $U_L$ and $U_R$ have no overlap. We now argue that this means that $U_{L,R}$ are \emph{separately} symmetries, at least in the ground state subspace (however, if $U = U_L U_R$ holds even for excited states --as is the case for strong zero modes-- then the following argument applies to the full Hamiltonian). Decompose the Hamiltonian $H = H_L + H_R$ where $H_L$ has no overlap with $U_R$ and $H_R$ has no overlap with $U_L$. This is possible due to the locality of the Hamiltonian. (Note that $H_L$ will have overlap with $H_R$.) Due to the tensor product structure of the symmetry, we can also choose $H_L$ and $H_R$ such that $U$ is a symmetry of each individually. This means $0 = [U,H_L] = [U_LU_R,H_L] = [U_L,H_L] U_R$. Since $U_R$ is invertible, this means $[U_L,H_L] = 0$. The fact that $U_L$ has no overlap with $H_R$ also means $[U_L,H_R]= 0$. Hence $[U_L,H] = [U_L,H_L + H_R] = 0$. Similarly $[U_R,H] = 0$.

\textbf{Projective representation on the edge} \hspace{5pt} The previous paragraph showed that bulk symmetries $U,V \in G$ define edge symmetries $U_L,V_L,U_R,V_R$. We now discuss what relations hold for these edge operators, working in the bosonic setting --later we mention what changes in the fermionic case. Suppose, for example, that the original symmetries $U$ and $V$ are commutative, i.e. $UVU^{-1}V^{-1} =1$, then $U_LV_L = e^{i\alpha} V_L U_L$. To see this, note that 
\begin{align}
1 &= (U_LU_R)(V_LV_R)\left(U^{-1}_L U^{-1}_R\right)\left(V^{-1}_LV^{-1}_R\right) \\
&= \left(U_L V_L U_L^{-1} V_L^{-1}\right) \; \left(U_R V_R U_R^{-1} V_R^{-1}\right)\notag
\end{align}
Since the two factors act on disjoint regions, each must be proportional to the identity: $U_L V_L U_L^{-1} V_L^{-1} = e^{i\alpha}$. This proves the above claim. More generally, any group relation that holds in $G$, also holds for the edge symmetries \emph{up to a phase factor}. This means the edge transforms under a \emph{projective} representation of $G$.

\textbf{Gauge symmetry and classes} \hspace{5pt} The phase factors of such a projective representation can have an arbitrariness to them. The defining relationship $U = U_L U_R$ is invariant under the gauge transformation $U_L \to e^{i\beta}U_L$ and $U_R \to e^{-i\beta} U_R$. However, the above $e^{i\alpha}$ is unchanged. We say the phase defined by $U_L V_L U_L^{-1} V_L^{-1}$ is gauge invariant. On the other hand, if $U^2 = 1$, then $U_L^2 = e^{i\gamma}$ which transforms under the previous gauge transformation as $U_L^2 = e^{i(2\beta + \gamma)}$. In particular, one can (partially) fix the gauge of $U_L$ by choosing $U_L^2 = 1$. To each projective representation, one can associate its gauge-invariant phase factors. We say two projective representations belong to the same class if these factors are the same. For example, all half-integer projective representations of $SO(3)$ belong to the same class. The set of these classes itself forms a group (for example one can add two classes by multiplying their phase factors) which is mathematically denoted by $H^2(G;U(1))$ and is called the second group cohomology group with coefficients in $U(1)$. For example, $H^2_\textrm{grp}(SO(3);U(1)) = \mathbb Z_2$, corresponding to the two distinct classes of half-integer and integer spins. In case $G$ is finite, it is also referred to as the Schur multiplier of $G$.

\textbf{Topological invariants and protected edge modes} \hspace{5pt} The above shows that to each gapped symmetry-preserving Hamiltonian, we can associate a list of phase factors to its edges. If one has two different Hamiltonians, each with its own set of phase factors (i.e. each is associated to a class of projective representations), then if these phase factors cannot be smoothly deformed into one another, these Hamiltonians must be in distinct phases. This happens if these phase factors can only take discrete values. Consider for example $G = \mathbb Z_2 \times \mathbb Z_2$ generated by $U$ and $V$. We have already encountered the invariant $U_L V_L = e^{i\alpha} V_L U_L$. Since also $U^2=1$, then $U_L^2$ is a phase factor and hence $[U_L^2,V_L] = 0$. This means $e^{i2\alpha} = 1$, such that the projective representations of $G = \mathbb Z_2 \times \mathbb Z_2$ are labeled by $U_L V_L = \pm V_L U_L$. Such a discrete invariant cannot change smoothly and thus labels distinct phases. Note that a non-trivial projective representation always has a dimension $>1$ (otherwise everything would trivially commute). In this way non-trivial phase factors are also linked to degenerate edge modes. More concretely, a $d$-dimensional projective representation protects a $d$-dimensional edge mode.

Not all distinct classes of projective representations define different phases. For example, the projective representations of $G= \mathbb Z \times \mathbb Z$ are characterized by a continuous phase $U_L V_L = e^{i\alpha} V_L U_L$. In other words, the distinct classes of projective representations are labeled by $H^2_\textrm{grp}(\mathbb Z \times \mathbb Z; U(1)) = U(1)$: there are infinitely many, but they are all smoothly connected. However, a finite-dimensional unit cell is symmetric with respect to a finite group $G$ or a compact Lie group $G$, in which case $H^2_\textrm{grp}(G;U(1))$ is discrete\footnote{If $G$ is a compact Lie group, then $H^2_\textrm{grp}(G;U(1))\cong H^2_\textrm{sing}(G;\mathbb Z)$ which is well-known to be $\cong \mathbb Z^\beta \oplus T$ with $\beta \in \mathbb N$ and $T$ finite.}. So for the case of finite-dimensional on-site Hilbert spaces, the classes of projective representations are characterized by \emph{discrete} invariants, i.e. they label distinct SPT phases with protected edge modes.

\textbf{Anti-unitary symmetries} \hspace{5pt} A similar procedure works for an anti-unitary symmetry $T = UK$, where $U$ is an on-site symmetry and $K$ is complex conjugation defined in a tensor product basis. If one chooses a basis for the low-energy degrees of freedom (necessarily living on the edge since the bulk is gapped) which factorizes between left and right, then one can define a new notion of complex conjugation, $\tilde K$, with respect to this factorized basis. If we restrict ourselves to these basis states, the same argument goes through as before, i.e. the symmetry will effectively act as $T = U_L U_R$. Allowing for phase factors and superpositions, the expression becomes $T = U_L U_R \tilde K$ in the low-energy subspace.

If the original symmetry satisfies $T^2=1$, then
\begin{align}
1 &= T \; ( U_L U_R \tilde K ) \notag \\
&= T U_LT^2 U_R T^2 \tilde K \notag \\
&= T U_L T T U_R T U_L U_R \tilde K^2 \notag \\
&= ( \overline U_L U_L) (\overline U_R U_R) \quad \textrm{ where } \overline{\mathcal O} := T \mathcal O T
\end{align}
Since the two factors act on disjoint regions, $\overline U_L U_L = e^{i\kappa}$. Note that this phase factor is invariant under $U \to e^{i\alpha} U$. Moreover we see that $U_L^{-1} = e^{-i\alpha} \overline U_L$, and since any operator commutes with its inverse, we have that $U_L$ and $\overline U_L$ commute. Hence the product $\overline U_L U_L$ must be real. We conclude that the projective representations of $T^2=1$ are labeled by $U_L \overline U_L = \overline U_L U_L = \pm 1$. Alternatively, one could have defined the invariant $U_L \tilde K U_L\tilde K = \pm 1$, and in fact for bosonic systems $(U_L T)^2 = (U_L \tilde K)^2$ (which can be proven using $T = T^{-1} = \tilde K U_R^{-1} U_L^{-1}$) so the choice is irrelevant. The latter choice might seem more natural, since $U_L \tilde K$ can be said to be an anti-unitary operator living on the left edge, but the fermionic case (which we address soon) shows that the other invariant is preferable.

To confirm that this invariant is independent of our choice of (factorized) basis, note that any other choice leads to a complex conjugation $\tilde K' = W_L W_R \tilde K W_R^{-1} W_L^{-1}$. Each effective complex conjugation, $\tilde K$ and $\tilde K'$, leads to a fractionalization $T = U_L U_R \tilde K = V_L V_R \tilde K'$. Substituting the above expression for $\tilde K'$, one obtains $U_L = V_L W_L \tilde K W_L^{-1} \tilde K$ up to a phase factor which does not affect the argument. Using this one can indeed straightforwardly show that $(U_LT)^2= (V_LT)^2$, again using the trick that $T = T^{-1}$.

\textbf{What changes for fermions} \hspace{5pt} So far we have used the fact that if $U_L$ and $U_R$ act on disjoint regions, then they commute. This clearly need not be the case for fermionic systems. This means that for each symmetry we can now have an extra phase factor: $U_L U_R = \pm U_R U_L$. Equivalently, this encodes whether $U_L$ is bosonic or fermionic, i.e $U_L P = \pm P U_L$, where $P$ is fermionic parity symmetry. A (projective) representation with this extra structure is called graded\cite{Fidkowski11}.

There is an important subtlety. In order to have a well-defined symmetry fractionalization of an anti-unitary symmetry, $T=U_LU_R \tilde K$, it is important that $\tilde K$ is chosen with respect to basis that factorizes over the edges. If this can be done, then the above proof directly applies to show the gauge invariance of $U_L \overline U_L$, even if $U_L$ is fermionic. However, fermionic chains can have a non-local fermionic mode that is spread out over both edges and hence such a basis does not exist. The best one can do is a decomposition $\mathcal H = \left( \mathcal H_L \otimes \mathcal H_R \right) \oplus \mathcal H_\textrm{non-local}$, where $\dim \mathcal H_\textrm{non-local} = 0,2$. This corresponds to respectively having an even or odd number of Majorana modes per edge. The definition of $\tilde K$ then depends on the basis one chooses in $\mathcal H_\textrm{non-local}$, which can possibly change the value of $U_L \overline U_L$. This simply means the anti-unitary symmetry protects the non-local mode (e.g. this is the case for the Kitaev chain, which is dual to the statement that the degeneracy of the Ising chain is protected by the spontaneously broken $PT$ symmetry). Despite $U_L \overline U_L$ not being gauge-invariant in that case, one can still use it to label distinct phases, even if one does \emph{not} make consistent gauge choices --this will be illustrated in the example of the $\alpha$-chain which we soon compute. Nevertheless, if one so prefers, one can consistently fix the gauge by requiring that the non-local basis vectors are chosen to be an eigenstate of $P_L$ (where $P$ is fermionic parity symmetry). Equivalently this means $\tilde K P_L \tilde K = P_L^{-1}$. Note that this condition on $\tilde K$ is independent of the gauge choice for $P_L$.

One might wonder how what changes if we switch between the two possible gauge choices. To this purpose, we can label the gauge by $\beta$, i.e. $\tilde K P_L \tilde K P_L = (-1)^\beta$. One can straight-forwardly prove that if $P_L P_R = P_R P_L$, then $\beta = 0$, confirming that the subtlety of fixing $\beta$ only arises in the presence of a non-local mode. In the latter case, one can show that $\tilde K P_R \tilde K P_R = (-1)^{\beta + 1}$, i.e. fixing this gauge is equivalent to choosing an edge, matching the fact that after a Jordan-Wigner transformation (which also chooses an edge) one obtains a spin chain where these subtleties do not arise. Suppose now that $T = U_L U_R \tilde K$ in a gauge labeled by $\beta$, then one can change the gauge by choosing $\tilde K' = P_L P_R \tilde K$. One can show that the new fractionalization, $T = V_L V_R \tilde K'$, satisfies $V_L \overline V_L = (-1)^{\beta + af} U_L \overline U_L$, where $a$ (resp. $f$) denotes whether $P_L$ (resp. $U_L$) is fermionic. Similarly, the same identity holds for the right-hand side if we replace $\beta \to \beta + a$. (Useful intermediate results to prove this, are $P_L T P_L T = (-1)^{\beta +f }$ and $P_L U_L = (-1)^{(a+1)f} U_L P_L$, which both straightforwardly follow from the trick of rewriting $P_L = P P_R^{-1}$ and $T = T^{-1} = \tilde K U_R^{-1} U_L^{-1}$.)

Another subtlety is that instead of the invariant $(U_LT)^2$ one could consider $(U_L \tilde K)^2$. However, one can show that $(U_LT)^2 = \pm (U_L \tilde K)^2$, where the sign corresponds to $U_L$ being bosonic (plus) or fermionic (minus). Hence if one is merely interested in counting and distinguishing phases, the choice is irrelevant. However, in section \ref{sec:alpha} we have argued that the former choice is more natural in terms of the physics. For example, it leads one to the conclusion that the $2$-chain is protected by $PT$ on the left-hand side, which is indeed given substance by the Jordan-Wigner transformation (with its string starting at the left end) mapping the $2$-chain to a spin chain protected by $PT$ (and not $T$).

\textbf{Symmetry fractionalization of the $\bm \alpha$-chain} \hspace{5pt} The $\alpha$-chain is a fermionic system with an anti-unitary symmetry $T = K$. From the above discussion, one can make an educated guess about the number of phases it has: there is an invariant for whether or not the fractionalization of $P$ is fermionic (i.e. there are an odd number of Majorana modes per edge) and then two invariants for whether or not $T$ protects something on the left or right. In summary we are interested in obtaining for each $\alpha$-chain the following phase factors (where $T = U_L U_R \tilde K$):
\begin{align} \label{def:abc}
P_L P_R &= (-1)^a P_R P_L \\
T U_L TU_L = \overline U_L U_L &= (-1)^b \\
T U_R TU_R = \overline U_R U_R  &= (-1)^c
\end{align}
Note that if one is given $b$, then the invariant $c$ is equivalent to the information of whether or not $U_{L,R}$ is fermionic. Indeed: $1=T^2 = TU_L U_R \tilde K = (T U_L T)(T U_R T) U_L U_R $, hence the fractionalization being bosonic or fermionic is equivalent to $(U_LT)^2$ having, respectively, the same or opposite sign as $(U_R T)^2$. One can rephrase this as $U_L U_R = (-1)^{a+b} U_R U_L$, and also $U_L P = (-1)^{a+b} P \; U_L$. Note that as discussed above, the values of $b$ and $c$ depend on the choice of complex conjugation in case of a non-local fermion (i.e. $a=1$). One can encode this choice in $\beta = 0,1$ where $P_L \tilde K P_L \tilde K = (-1)^\beta$. Nevertheless, we will see $a,b,c$ successfully distinguish all eight phases \emph{even} if one mixes choices of $\beta$.

A priori one might also expect $PT$ to give extra invariants. However we now show that its fractionalization is fixed by the above information. If we write $PT = V_L V_R \tilde K$, then
\begin{equation}
(PT) V_L (PT) V_L = 
\left\{
\begin{array}{rl}
T U_R T U_R & \quad \textrm{ if }a=0\\
(-1)^{\beta + 1} T U_L T U_L & \quad \textrm{ if }a=1 
\end{array}
\right.
\end{equation}
This is straight-forward to derive. Firstly note that $V_L = P_L U_L$ (up to an irrelevant sign), hence
\begin{align}
(PT) V_L (PT) V_L &= PT P_L U_L PT P_L U_L \notag \\
&= P T P_L T T U_L T P P_L U_L \notag  \\
&= (-1)^{\beta + b+c + a(b+c)} P P_L^{-1} T U_L T U_L P P_L \notag \\
&= (-1)^{\beta + b+c+ a(b+c)+a} T U_L T U_L
\end{align}
where we have used that $P_L T P_L T = (-1)^{\beta + b+c}$ and $U_L P P_L = (-1)^{a(b+c)} P P_L U_L$.

We now explicitly derive the expressions for $P_{L,R}$ and $U_{L,R}$ for the $\alpha$-chain (where for notational convenience we choose $\alpha$ positive). One may easily ascertain that up to an irrelevant sign
\begin{equation} \label{PL}
P_L = \prod_{1 \leq n\leq \alpha} \gamma_n \qquad P_R = i^\alpha \prod_{0 \leq n< \alpha} \tilde \gamma_{N-n}
\end{equation}
This is a direct consequence of $P = i^N \prod \tilde \gamma_n \gamma_n$ and the fact that for all $1\leq n \leq N-\alpha$, in the ground state subspace $\tilde \gamma_n \gamma_{n+\alpha } =i$. To factorize the low-energy Hilbert space made up by these modes as much as possible onto the edges, let us define 
\begin{equation}
\left\{
\begin{array}{lll}
c^L_1 = \frac{1}{2} \left( \gamma_1 + i\gamma_2 \right) & & c^R_1 = \frac{1}{2} \left( \tilde \gamma_{N-1} + i\tilde \gamma_N \right) \\
\quad \; \; \vdots &&\quad \; \; \vdots \\
c^L_a = \frac{1}{2} \left( \gamma_{2a-1} + i\gamma_{2a} \right) & & c^R_a = \frac{1}{2} \left( \tilde \gamma_{N-2a+1} + i\tilde \gamma_{N-2(a-1)} \right)
\end{array} \right.
\end{equation}
where $a=\lfloor \alpha/2 \rfloor$. If $\alpha$ is odd we have the extra non-local mode $c = \frac{1}{2} \left( \gamma_\alpha + i\tilde \gamma_{N-\alpha+1} \right) $. We now define $\tilde K$ as complex conjugation in the basis of these fermionic modes. Equivalently:
\begin{equation}
\quad \tilde K \stackrel{\Scale[0.5]{(\sim)}}{\gamma}_n \tilde K = (-1)^{n+1} \stackrel{\Scale[0.5]{(\sim)}}{\gamma}_n  \label{def:K}
\end{equation}
One can ascertain that in this gauge we have $P_L \tilde K P_L \tilde K = 1$, i.e. $\beta = 0$ (the other gauge would correspond to changing $(-1)^{n+1} \to (-1)^n$). Comparing Eq.~\eqref{def:K} to the action of $T$, i.e. $T \gamma_n T = \gamma_n$ and $T \tilde \gamma_n T = - \tilde \gamma_n$, we see that
\begin{equation}\label{UL}
U_L = \prod_{1\leq \textrm{even } n \leq \alpha} \gamma_n \qquad U_R = \prod_{0 \leq \textrm{odd } n< \alpha} \tilde \gamma_{N-n}
\end{equation}

The above explicit symmetry fractionalizations allow us to read off the invariants $a,b,c$, as summarized in Table.~\eqref{table:abc}. As an example, consider $\alpha = 3$ such that $U_L = \gamma_2$. Then $U_L \overline U_L = \gamma_2 (-\gamma_2) = -1$, hence $b=1$.

\begin{table}[h]
	\begin{tabular}{c||c|cc|cc}
		$\alpha$ & $ \quad a\quad $ & $\quad b \quad$ &$\qquad$ & $\quad c\quad$ &$\qquad$  \\ 
		\hline $0$ & $0$ & $0$ && $0$ & \\
		$1$ & $1$ & $0$ && ($0$) & [$1$] \\
		$2$ & $0$ & $0$ && $1$& \\
		$3$ & $1$ & $(0)$ & [$1$]& $1$ & \\
		$4$ & $0$ & $1$ && $1$& \\
		$5$ & $1$ & $1$ && ($1$) &[$0$] \\
		$6$ & $0$ & $1$ && $0$& \\
		$7$ & $1$ &($1$) &[$0$] & $0$ & \\
	\end{tabular}
	\caption{The phase factors characterizing the symmetry fractionalization of $P$ and $T$ as defined in Eq.~\eqref{def:abc} and derived from Eqs.~\eqref{PL} and Eq.~\eqref{UL}. If the result depends on the gauge choice $P_L \tilde K P_L \tilde K = (-1)^\beta$, we show it in parentheses. In that case, the value in round (square) brackets corresponds to $\beta = 0$ ($\beta =1$). Note that these three columns correspond to the first two columns in Table~\eqref{table:alpha_symfrac}.}
	\label{table:abc}
\end{table}

From our earlier discussion (and characterization) of how the symmetry fractionalization of $T$ depends on the choice of basis, we can also directly obtain the values for the gauge choice $P_R \tilde K P_R \tilde K = 1$ (i.e. $\beta = 1$ if $a=1$). When $b$ or $c$ depend on this choice, we show it in parentheses, where value in round (square) brackets corresponds to $\beta = 0$ ($\beta =1$). Note that one can also directly calculate it in the basis where $P_R \tilde K P_R \tilde K = 1$ by redefining $\tilde K\to P_L P_R \tilde K$, in which case the sign in Eq.~\eqref{def:K} changes from $(-1)^{n+1}$ to $(-1)^n$. For example, $U_L$ is now given by the product of \emph{odd} Majorana modes instead of even ones.

This information is represented in the main text in Table~\eqref{table:alpha_symfrac}. There we have inserted an extra column specifying the symmetry fractionalization of $PT$, which can be derived from that of $P$ and $T$ as mentioned before. Note that changing the gauge choice is equivalent to swapping the $T \leftrightarrow PT$ and `left' $\leftrightarrow$ `right' in Table~\eqref{table:alpha_symfrac}. This allows one to directly see which values are gauge-independent.

\end{document}